\documentclass[useAMS,usenatbib]{mn2e}
\usepackage{graphicx}
\usepackage{array}
\usepackage{amssymb}
\usepackage{txfonts}
\usepackage{textcomp}
\usepackage{mabib}
\usepackage{natbib}
\begin{document}

\title[Pulsations in B stars of the Magellanic Clouds]  {Testing the effects of opacity and the chemical mixture on the excitation of pulsations in B stars of the Magellanic Clouds}

\author[S. Salmon et al.]
  {S.~Salmon,$^1$\thanks{E-mail:sebastien.salmon@ulg.ac.be}
  J.~Montalb\'{a}n,$^1$ T.~Morel,$^1$ A.~Miglio,$^{1,2}$ M-A.~Dupret$^1$ and A.~Noels$^1$\\
  $^1$Institut d'Astrophysique et de G\'eophysique de l'Universit\'e de Li\`ege, All\'ee du 6 Ao\^ut 17, 4000 Li\`ege, Belgium\\
  $^2$School of Physics and Astonomy, University of Birmingham, Edgbaston, Birmingham B15 2TT, UK}

\date{Released 2002 Xxxxx XX}

\pagerange{\pageref{firstpage}--\pageref{lastpage}} \pubyear{2002}

\def\LaTeX{L\kern-.36em\raise.3ex\hbox{a}\kern-.15em
    T\kern-.1667em\lower.7ex\hbox{E}\kern-.125emX}

\maketitle

\label{firstpage}

\begin{abstract}
The B-type pulsators known as $\beta$~Cephei and Slowly Pulsating B (SPB) stars present pulsations driven by the $\kappa$ mechanism, which operates thanks to an opacity bump due to the iron group elements. In low-metallicity environments such as the Magellanic Clouds, $\beta$ Cep and SPB pulsations are not expected. Nevertheless, recent observations show evidence for the presence of B-type pulsator candidates in both galaxies.\\ We seek an explanation for the excitation of $\beta$~Cep and SPB modes in those galaxies by examining basic input physics in stellar modelling: \emph{i)} the specific metal mixture of B-type stars in the Magellanic Clouds; \emph{ii)} the role of a potential underestimation of stellar opacities.\\
We first derive the present-day chemical mixtures of B-type stars in the Magellanic Clouds. Then, we compute stellar models for that metal mixture and perform a non-adiabatic analysis of these models. In a second approach, we simulate parametric enhancements of stellar opacities due to different iron group elements. We then study their effects in models of B stars and their stability.\\
We find that adopting a representative chemical mixture of B stars in the Small Magellanic Cloud  cannot explain the presence of B-type pulsators there. An increase of the opacity in the region of the iron-group bump could drive B-type pulsations, but only if this increase occurs at the temperature corresponding to the maximum contribution of Ni to this opacity bump. We recommend an accurate computation of Ni opacity to understand B-type pulsators in the Small Magellanic Cloud, as well as the frequency domain observed in some Galactic hybrid $\beta$~Cep-SPB stars.
\end{abstract}


\section{Introduction}
\label{intro}

$\beta$ Cephei and SPB pulsators are main sequence (hereafter MS) B stars presenting self-driven oscillations. The former present low-order acoustic and gravity modes (respectively p and g modes), while the latter show high-order g modes. $\beta$ Cep stars have masses from 7 to 20 M$_{\odot}$, with typical periods between 2 and 8 hours \citep{handler}. SPB masses range from 2.5 to 8  M$_{\odot}$, with typical periods between 1 and 3 days \citep{waelkens}. These oscillations are excited through the $\kappa$ mechanism, activated by the presence of an opacity bump at $T \approx 200,000$~K due to the iron group elements (\citealp[see][]{dziembowski92,cox}; \citealt*{dziembowski93}).

The opacity profile, shaped by the iron group elements, determines the limits of the instability strip of B-type pulsators and the frequency domain of their oscillations. Individual abundances of the iron group elements (Fe, Ni, Cr and Mn) and their distribution inside the star, as well as their contribution to stellar opacity are essential to understanding the seismic properties of B-type pulsators. Indeed, the nature of $\beta$~Cep pulsations could only be understood once new opacity computations by OPAL \citep{rogers96}  and OP \citep{seaton94},  including  the contribution of a huge number of bound-bound transitions from the iron group elements, were available. These opacity computations were then able to account for most of the observational data for $\beta$ Cep and SPB pulsators available at the time \citep[e.g.][]{pami99} and successfully fit asteroseismic observables of $\beta$ Cep stars such as 16~Lac \citep{Jerki96,Thoul03}, HD 129929 \citep{Aerts03,Dupret04} or $\beta$~CMa \citep{Mazumdar06}.

With progress in observational techniques, more and more frequencies were observed in $\beta$~Cep and SPB stars while objects presenting both low-order p and high-order g modes (hybrid pulsators) were found. As a consequence, new discrepancies between observational data and theoretical predictions appeared. Models were not able to account for oscillation spectra of hybrid pulsators (\citealp*[e.g. 12~Lac and $\nu$~Eri:][]{dziembo04}; \citealt{Ausseloos04}), that is for the simultaneous detection of modes spanning a large frequency domain, including SPB and $\beta$ Cep pulsations.

In the last decade, two events, the update of the solar mixture (\citealp*[][hereafter AGS05]{asplund}) and new opacity computations by OP \citep{badnell}, led to important consequences for the properties of B-type pulsators. The revision of the solar mixture decreased the abundances of C, N, O and Ne by $\sim 30$\%  with respect to  the \textquotedblleft standard\textquotedblright solar mixture \citep[][hereafter GN93]{gn93}, while the Fe one did not change. For a fixed metallicity (Z), the mass fraction of Fe is then 25\% larger in the new mixture than in GN93. On the other hand, the OP team improved opacity computations \citep{badnell} by including new inner-shell atomic data for 13 elements and by complementing previous data with new inter-combination lines and photo-ionization cross-sections. In particular, this led to an 18\% increase of the Rosseland mean opacity in the region of the iron opacity bump. As a result, for the structure of typical $\beta$~Cep stars, the iron opacity bump is modified by up to $\sim$~30\% (see Fig. 4 of \citealt{badnell}, see also \citealt*{mmma}) compared to what would be obtained with the commonly used OPAL opacities \citep{rogers96}.

\citet{mmma} studied the effects of the solar mixture (GN93 {\it vs} AGS05) and the opacity data (OPAL {\it vs} OP) on mode stability in MS B stars and found that the most favourable case for the excitation of $\beta$~Cep and SPB modes was obtained by combining the AGS05 mixture with the new OP data. The relative increase, at a given Z, of the mass fraction of iron group elements in the AGS05 mixture yields wider instability bands than does the GN93 mixture. The hot wing of the opacity bump in OP data also extends the blue edge (hotter limit) of the instability strip in comparison with OPAL. Finally, the OP opacity profile leads to a larger frequency domain of excited modes and to a  greater number of hybrid pulsators, as SPB modes are more easily excited (\citealp*[see also][]{miglio}; \citealp{pami07,pami08}).

The combination of the AGS05 mixture and the updated OP opacities has now been used in several studies. \cite{dziembowski4} obtained a larger frequency domain of excited modes in modelling $\nu$ Eri while \cite{Briquet07} had promising results for the $\beta$ Cep star $\theta$ Oph. However, the lowest frequency modes detected in $\nu$~Eri still cannot be theoretically reproduced \citep{dziembowski4}. Additional seismic studies show that OPAL is better at reproducing the non-adiabatic properties of $\beta$ Cep modes in $\theta$ Oph and $\nu$~Eri \citep{jagoda09,jagoda10}, whereas OP is preferable for an SPB mode in $\nu$ Eri \citep{jagoda10}. The excitation of the lowest frequencies of 12~Lac are however still not explained by theoretical models \citep{dziembowski4,Desmet09}.

The disagreement between observations and theoretical predictions based on updated input physics is even stronger when low-metallicity environments are considered. Theoretical computations with AGS05 and OP \citep{mmma} predict SPBs and $\beta$~Cep stars down to respectively $Z$~=~0.005 and $Z$~=~0.01. Therefore, these pulsators should not be present in the Small Magellanic Cloud (SMC) and only rarely in the Large Magellanic Cloud (LMC), galaxies for which \citet{buchler}, from a study of Cepheid properties, derived average metallicities of $Z$~=~0.0027 for the SMC and $Z$~=~0.0046 for the LMC. Nevertheless, \citet{Kola} announced the discovery of 151 candidates (92 $\beta$ Cep stars and 59 SPBs) in the LMC, and 17 (6 $\beta$ Cep stars and 11 SPBs) in the SMC. Moreover, \citet{sarro} found 63 candidates (48 $\beta$ Cep stars and 15 SPBs) in the LMC and 43 (23 $\beta$ Cep stars and 20 SPBs) in the SMC. B-type pulsators have also been detected in SMC stellar clusters: 29 SPB candidates in  NGC 371  \citep{karoff}, and 1 $\beta$ Cep candidate and 8 SPB candidates in NGC 330 \citep{diago}. 

All that calls into question some of the input physics used in modelling opacity profiles and consequently the stability of pulsation modes. It is generally assumed that for population~I stars, the distribution of element abundances in the stellar matter is the same as in the Sun, although scaled to their metallicity ($Z$). However, as shown by \citet{mmma} and \citet*{montalbanliege}, changing the pattern of element abundances (even with the same value of $Z$) can have important implications for the stability analysis of B-type pulsators. Hence, we wonder whether the solar mixture is representative of MS B stars in the MCs. 

The problem could also arise from opacity calculations which may underestimate contributions in some particular temperature and density domain. This possibility has already been invoked to solve old and recent disagreements between theoretical and observational seismic properties of MS pulsators \citep[e.g.][]{simon,montalbansun,bahcall,montalban44tau,dziembowski4}, as well as of evolved  B-type pulsators \citep{jeffrey}. In particular, \citet{jeffrey} showed that increasing the opacity due to Ni could help with fitting the observational subdwarf B (sdB) instability strip. Fortunately, these remaining problems have motivated different teams to undertake experiments trying to reproduce stellar plasma conditions in a laboratory to measure monochromatic opacities \citep[e.g.][]{TC3}. The first comparisons with calculations made by OP and other plasma opacity codes, including attempts to improve theoretical calculations, reveal discrepancies, in particular for the Fe and Ni monochromatic opacities (Turck-Chi{\`e}ze et al. 2011\nocite{TC1}; Gilles et al. 2011\nocite{Gilles112}, and references therein), suggesting that there is still room for improvement in opacity data, with important consequences for B-type pulsators.

Although we are aware that non-standard processes might have some effects on B star pulsations, our aim in this paper is to address the role of standard input physics on the driving of pulsations in MS B stars. We begin with detailing the chemical mixtures representative of B stars in the MCs in the following section. In the next section, we then consider the pulsation stability of B star models computed with the SMC chemical mixture. In Sect. \ref{incopacity}, we analyse the effects of an {\it ad hoc} local change in the opacity, to identify the changes that could explain the presence of B-type pulsators in the SMC. A discussion and a conclusion end the paper.

\section{Chemical mixture of young stellar objects in the SMC}

\label{sect_abundances}
\label{smcabundances}

The detailed chemical mixture and opacity data are two basic factors that can affect the driving of pulsations in $\beta$ Cep stars and SPBs. The Magellanic Clouds are a very convenient place to observe not only the effects due to a different chemical mixture but also those coming from a low metallicity. The results of \citet{mmma} predict a limited number of SPBs and no $\beta$~Cep stars in the MCs. These results are challenged by recent detection of a significant number of B-type pulsator candidates in the MCs, and particularly in the SMC, the most metal-poor Cloud.

\begin{table*}
\centering
\caption{Adopted logarithmic metal abundances in the SMC (on the scale in which $\log \epsilon$[H]=12 and rounded to the nearest 0.05 dex), along with the source of the data. We consider the mean value (weighted by the number of lines) when abundances are given for different ions of the same element. The scaled solar mixture taking Fe as a reference is given for comparison \citep[solar values from][]{AGS09}. The last row gives the resulting metallicity \citep[assuming that the abundances of the elements not listed is 0.75 dex below the values of][]{AGS09}.}

\begin{tabular}{lcclc} \\\hline
Element  & SMC &  Scaled solar & Source & Reference\\
 & &  (Asplund -- 0.75 dex) & &\\
\hline 
C  &  7.50  &  7.68 & B stars & 1, 2\\  
N  &  6.55  &  7.08 & B stars/H II regions & 2, 3, 4, 5, 6, 7 \\  
O  &  8.00  &  7.94 & B stars & 1, 2, 3, 8 \\  
Ne &  7.25  &  7.18 & H II regions & 4, 5, 6, 7, 9\\   
Na &  5.60  &  5.49 & Cool supergiants & 10, 11, 12, 13, 14\\   
Mg &  6.75  &  6.85 & B stars & 1, 2, 3, 8 \\   
Al &  5.60  &  5.70 & B stars & 1\\  
Si &  6.80  &  6.76 & B stars & 1, 2, 3, 8 \\   
S  &  6.40  &  6.37 & B stars & 2\\   
Ar &  5.80  &  5.65 & H II regions & 4, 5, 6, 7, 9\\    
Ca &  5.55  &  5.59 & Cool supergiants & 10, 11, 12, 13, 14, 15\\   
Sc &  2.25  &  2.40 & Cool supergiants & 10, 11, 12, 13, 14, 15\\   
Ti &  4.30  &  4.20 & Cool supergiants & 10, 11, 12, 13, 14, 15\\    
V  &  3.05  &  3.18 & Cool supergiants & 11, 12, 15\\    
Cr &  4.90  &  4.89 & Cool supergiants & 10, 11, 12, 14, 15\\    
Mn &  4.85  &  4.68 & Cool supergiants & 12, 15\\    
Fe &  6.75  &  6.75 & Cool supergiants & 10, 11, 12, 14, 15, 16\\   
Co &  4.00  &  4.24 & Cool supergiants & 11, 12, 13\\   
Ni &  5.30  &  5.47 & Cool supergiants & 10, 11, 12, 13, 15\\   
\hline 
{\it Z} ($\times$ 10$^{-3}$) & 2.36  & 2.41 & &\\ 
\hline 
\end{tabular}
\begin{flushleft}
Key to references (for meaning of acronyms see Table \ref{tab_acronyms}) --- [1] \citet{korn00}, [2] \citet{hunter05}, [3] \citet{hunter07}, [4] \citet{reyes99}, [5] \citet{vermeij02}, [6] \citet{kurt99}, [7] \citet{garnett99}, [8] \citet{trundle07}, [9] \citet{lebouteiller08}, [10] \citet{hill97}, [11] Hill (1999), [12] L98 for MARCS models and spectroscopic gravities, [13] GW99, [14] \citet{venn99}, [15] RB89, [16] \citet{romaniello08}.
\end{flushleft}
\label{tab_abundances_smc}
\end{table*}

As a first approach, we try to determine whether that source of discrepancy between theory and observation comes from the chemical mixture. For that purpose we have assembled for the SMC and the LMC a set of representative absolute abundance values for all elements significantly contributing to the opacity from a comprehensive and critical examination of literature data. As the central bar is thought to have efficiently mixed the Interstellar Medium (ISM) through gas streaming, large-scale spatial chemical inhomogeneities should not be a major issue in the case of young objects, as indeed confirmed by observations \citep[e.g.][]{hill97,andrievsky01}. The breakdown of these abundance data for all elements up to $Z$~=~26 are shown in Fig. \ref{fig_abundances_smc} and \ref{fig_abundances_lmc} for the SMC and the LMC, respectively. Some minor chemical species are not discussed because of the lack of measurements, but their contribution to the global metallicity is negligible.

\begin{table*}
\centering
\caption{Adopted logarithmic metal abundances in the LMC (on the scale in which $\log \epsilon$[H]=12 and rounded to the nearest 0.05 dex), along with the source of the data. We consider the mean value (weighted by the number of lines) when abundances are given for different ions of the same element. The scaled solar mixture taking Fe as a reference is given for comparison \citep[solar values from][]{AGS09}. The last row gives the resulting metallicity \citep[assuming that the abundances of the elements not listed is 0.35 dex below the values of][]{AGS09}.} 

\begin{tabular}{lcclc} \\\hline
Element  & LMC &  Scaled solar & Source & Reference\\
 & &  (Asplund -- 0.35 dex) & &\\
\hline 
C  &  8.00 &  8.08 & B stars & 1, 2, 3\\  
N  &  6.90 &  7.48 & B stars & 4 \\  
O  &  8.35 &  8.34 & B stars & 4 \\  
Ne &  7.70 &  7.58 & H II regions & 5, 6, 7, 8, 9\\   
Na &  5.90 &  5.89 & Cool supergiants & 10\\   
Mg &  7.05 &  7.25 & B stars & 4 \\   
Al &  5.85 &  6.10 & B stars & 1\\  
Si &  7.20 &  7.16 & B stars & 4 \\   
S  &  7.10 &  6.77 & Cool supergiants & 10, 11, 12\\   
Ar &  6.15 &  6.05 & H II regions & 5, 6, 7, 8, 9\\    
Ca &  6.10 &  5.99 & Cool supergiants & 10, 11\\   
Sc &  2.65 &  2.80 & Cool supergiants & 10, 11, 12\\   
Ti &  4.70 &  4.60 & Cool supergiants & 10, 11, 12\\    
V  &  3.80 &  3.58 & Cool supergiants & 11, 12\\    
Cr &  5.40 &  5.29 & Cool supergiants & 10, 11, 12\\    
Mn &  4.95 &  5.08 & Cool supergiants & 10, 11, 12\\    
Fe &  7.15 &  7.15 & B stars/Cool supergiants & 1, 2, 10, 11, 12, 13, 14\\   
Co &  4.65 &  4.64 & Cool supergiants & 10, 11\\   
Ni &  5.85 &  5.87 & Cool supergiants & 10, 11, 12\\   
\hline 
{\it Z} ($\times$ 10$^{-3}$) & 5.96  & 6.05 & &\\ 
\hline 
\end{tabular}
\begin{flushleft}
Key to references --- [1] \citet{korn00}, [2] \citet{korn02}, [3] \citet{korn05}, [4] \citet{hunter08}, [5] \citet{reyes99}, [6] \citet{lebouteiller08}, [7] \citet{vermeij02}, [8] \citet{peimbert03}, [9] \citet{garnett99}, [10] A01, [11] L98 for MARCS models and spectroscopic gravities, [12] RB89, [13] \citet{trundle07}, [14] \citet{romaniello08}. 
\end{flushleft}
\label{tab_abundances_lmc} 
\end{table*}

The data are necessarily inhomogeneous and come from three independent sources: B stars, cool supergiants (note that they were chosen to cover a similar mass range as the B stars to ensure that they sample the same young population) and \mbox{H\,{\sc ii}} regions. All these tracers suffer to different extents from various limitations and sources of errors (e.g. neglect of departures from LTE and evolutionary effects for the stellar objects, depletion onto dust grains and conversion from ionic to elemental abundances for the ionised gas). Although preference is given to the abundances of B stars, in a few cases we feel that data from other sources are more reliable and decide to use them instead (e.g. Ne data from \mbox{H\,{\sc ii}} regions). For many elements, especially the heaviest ones, only data for cool stars are available. The NLTE corrections have been shown to be generally small and typically well within $\sim$~0.2 dex, either for C, O and Na in LMC F-type supergiants \citep{andrievsky01} or for O and Mg in SMC A-type supergiants \citep{venn99}. The LTE abundances of these metals in the evolved objects should therefore be relatively accurate, but this may not be true for other species \citep[e.g. N in A-type stars;][]{venn99}. More generally, we give a higher weight to NLTE results and to the most recent studies based on higher quality observational material and improved analysis techniques. We describe the available abundance data and the criteria used to select what we regard as the most probable present-day value for each element in Appendices A and B. They are summarized in Tables \ref{tab_abundances_smc} and \ref{tab_abundances_lmc} for the SMC and the LMC, respectively.

As we aim to concentrate on the more metal-poor SMC, we note that the Fe abundance presented in Table \ref{tab_abundances_smc} is in remarkable agreement with the average literature value for other young ($<$~1.5 Gyr) SMC objects, that is $<$~[Fe/H]~$>$~$\simeq$~--0.70~$\pm$~0.15 \nocite{piattia}\nocite{piattic}(\citealp[gathered from][]{carrera,parisi1,parisi2,depropis}; Piatti et al. 2007a,b; \citealt*{idiart}). We also note the good agreement between our $Z$ value and the mean value derived by \citet{buchler}. Indeed, we obtain $Z_{{\rm SMC}}$~=~0.00236 and estimate the error to be $\sigma_{{\rm SMC}}$~$\sim $~0.00079 ($Z_{{\rm LMC}}$~=~0.00596 and $\sigma_{{\rm LMC}}$~$\sim$~0.00232), while \citet{buchler} obtains $Z$~=~0.0027 for the SMC ($Z$~=~0.0046 for the LMC). We determine the error on metallicity by a propagation of the errors on individual element abundances. These errors are taken as the standard deviations of the different measurement distributions for each abundance (see Figures \ref{fig_abundances_smc} and \ref{fig_abundances_lmc}). When there are only a few data points for a given element, we use instead a fixed error of 0.15 dex, representative of typical errors given in the literature for abundances of MCs stars.

Finally, we have determined the opacity profiles of a given stellar structure by considering, respectively, the AGS05 scaled solar mixture and the SMC mixture. The differences between the two opacity profiles do not exceed 3~\%, while they are at most 6~\% when the LMC mixture is considered instead.

\section{Stability analysis of main sequence B-type stars from the SMC}
\label{mssa}

We study the stability properties of MS B stars in the SMC. With the stellar evolutionary code CLES \citep{scuflaire}, we have computed a grid of stellar evolution models with masses from 2.5 to 18 M$_{\odot}$, a metallicity $Z$~=~0.00236 and a He mass fraction fixed at $Y$~=~0.28. We have also computed some models with a lower abundance of He, $Y$~=~0.26, which only affects the mass of models characterized by a given effective temperature ($T_{{\rm eff}}$) and luminosity ($L$). Convection is treated according to the mixing-length theory \citep{bohm} with its parameter fixed at $\alpha_{{\rm MLT}}$~=~1.8. The surface boundary conditions follow Eddington's law ($T[\tau]$) for a grey atmosphere. The convective core overshoot parameter is set to 0.2 pressure scale heights. The OPAL equation of state is selected from \citet{rogers}. Nuclear reaction rates are from NACRE (Angulo et al. 1999)\nocite{angulo} with the revised $^{14}{\rm N}(p,\gamma)^{15}{\rm O}$ cross section from Formicola et al. (2004)\nocite{formicola}. Concerning opacity, we have built new opacity tables for the specific SMC B star metal mixture (hereafter B-SMC mixture) with the last updated version of the OP code \footnote{OP website: http://opacities.osc.edu/} \citep{badnell}. 
Finally, we have performed a non-adiabatic stability analysis of these models with the code MAD \citep{dupret}.

We do not find any SPB or $\beta$ Cep modes to be excited in models at the SMC metallicity, $Z$~=~0.00236. This is not surprising since opacity profiles with a scaled solar or a B-SMC mixture are very close and since results with solar mixtures already indicated no excitation at such a metallicity \citep[][]{mmma}. In addition, we have made grids of models with the same physics, but with a metallicity of $Z$~=~0.005 and $Z$~=~0.01. At $Z$~=~0.005, the B-SMC mixture only leads to excited SPB modes, while at $Z$~=~0.01 we find a similar instability domain as in the solar case for $\beta$ Cep stars and SPBs. 

\begin{figure}
\includegraphics[scale=0.38]{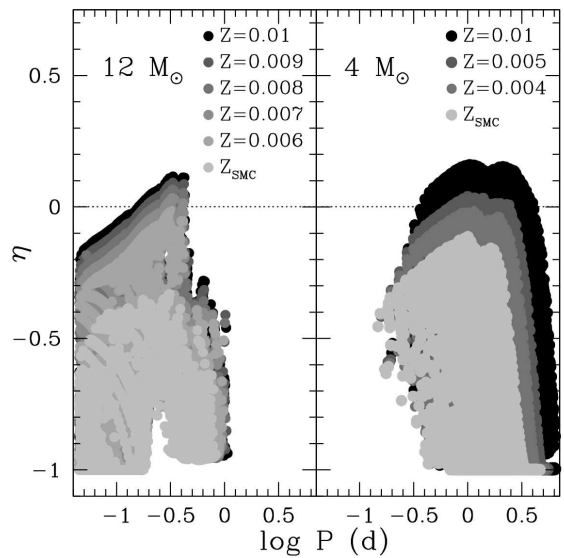}
\caption{Measurement of the instability of  $\beta$ Cep (left panel) low-order $\ell$~=~0, 1 and 2 p modes,  and SPB (right panel) high order $\ell$~=~1 and 2 g-modes as a function of the period of the mode, $\log P$ (in days). Models are taken along the MS evolution phase for 12 and 4 M$_{\odot}$  with the B-SMC mixture and for different values of $Z$.}
\label{bandinst}
\end{figure}

Finally, we derive the minimum values of $Z$ required to excite $\beta$ Cep and SPB modes in the SMC. We compute models of 4 and 12 M$_{\odot}$, respectively representative of an SPB and $\beta$ Cep star by varying the metallicity from $Z$~=~0.003 to 0.01 in steps of 0.001. In Fig.~\ref{bandinst}, we plot the normalized growth rate ($\eta$):
\begin{equation}
\eta=\frac{W}{\int_{0}^{R}\left|\frac{{\rm d}W}{dr}\right|dr},
\end{equation}  
where $R$ is the radius of the star. The represented $\eta$ values come from the eigenmodes of models from the Zero Age Main Sequence (ZAMS) to the Terminal Age Main Sequence (TAMS). If $\eta$ is positive, the work ($W$) developed by the whole star is positive, meaning energy is efficiently transferred to drive the mode. It turns out that $Z$~$>$~0.004 is required to excite SPB modes in the SMC, while $Z$~$\geq$~0.007 is needed for $\beta$ Cep modes.

In conclusion, the average B-SMC mixture and its corresponding metallicity ($Z_{{\rm SMC}}$~=~0.00236~$\pm$~0.00079) is marginally able to explain the presence of SPBs. But it is not able to explain the presence of $\beta$ Cep stars since it would imply that those stars have a metallicity significantly higher (at 6~$\sigma$) than the mean value for young SMC objects. 

\section{Constraining the iron-group opacity}
\label{incopacity}

\begin{figure}
\centering
\includegraphics[scale=0.38]{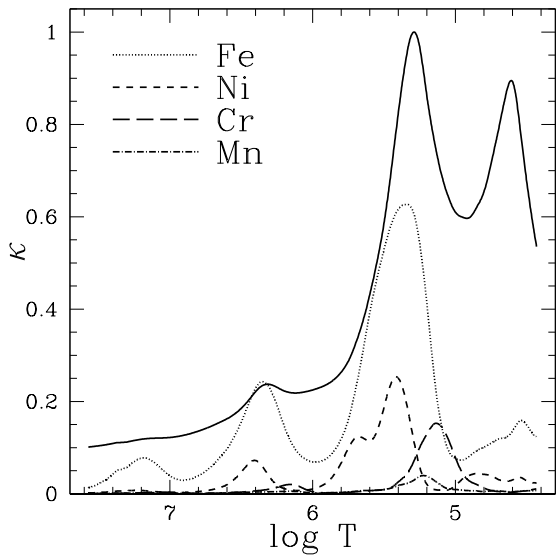}
\caption{The solid line represents the OP normalized opacity in a 12 M$_{\odot}$ model of a MS $\beta$ Cep star. The individual contributions of Fe, Ni, Cr and Mn are represented and expressed in percentage of the total opacity.}
\label{exampleofprofile}
\end{figure}

Already in the eighties, an underestimation of the opacity was suggested by \citet{simon}. He noticed that a revision of metal opacities by a factor of 2 to 3 could explain the pulsations in $\beta$ Cep stars. New opacity computations by OPAL \citep[][and references therein]{iglesias} confirmed it with a raise in the iron-group opacity by a factor of 3. More improvements to stellar opacities continued with the update of OPAL \citep{rogers96} and the work from OP \citep{seaton94,badnell}. Nowadays, some discrepancies remain on the predicted iron-group bump between OPAL and OP \citep{seaton}. Recent investigations on monochromatic opacities of Fe and Ni reveal some disagreements between current opacity tables and new computations (e.g. Gilles et al. 2011). Therefore, some room still remains for enhancing opacities. Opacity calculations rely on the equation of state of stellar plasmas and on the detailed atomic transitions (lines) of elements that constitute the plasma. Provided the equation of state is correct, opacities can only be underestimated. Indeed, various approaches are used to implement transition lines in opacity codes, in particular with somewhat different restrictions on the number of lines. We consider this caveat as a gateway to possible underestimations.

Hence we would like to constrain the opacity profile required to drive pulsations in MS B stars of the SMC. We proceed with an \emph{ad hoc} local change in the opacity by considering different cases where the increase could be due to poor estimation of the opacity of respectively Fe, Ni, Cr and Mn.

\begin{table}
\centering
\caption{Set of parameters used to simulate an increase of opacity by means of Gaussians centred on the contributions of Cr, Mn, Fe and Ni.}
\begin{tabular}{lclc} \\\hline
Element  & log T$_i$ &  $\sigma_i$ & A (in \%) \\
\hline 
Cr &  5.07  &  0.1875 & 5-10-20-30-50-100 \\    
Mn &  5.23  &  0.125 & 5/3-10/3-20/3-30/3-50/3-100/3\\    
Fe &  5.30  &  0.25 & 5-10-20-30-50-100\\   
Ni &  5.46  &  0.125 & 5-10-20-30-50-100\\   
\hline 
\end{tabular}

\label{param}
\end{table}

\begin{figure}
\centering
\includegraphics[scale=0.40]{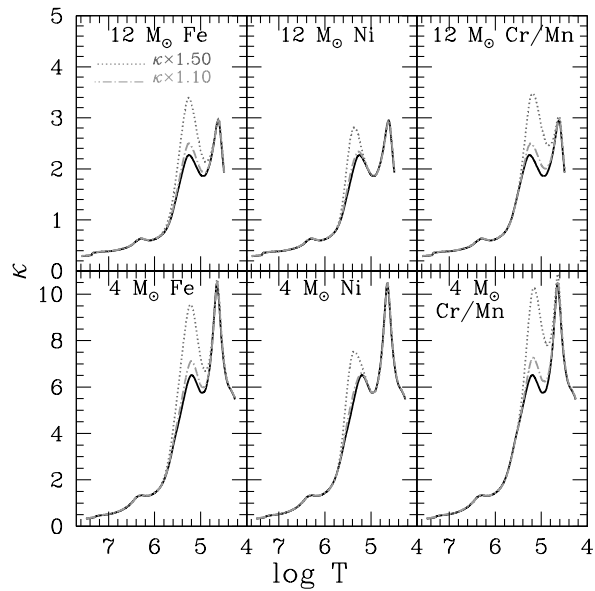}
\caption{OP opacity profiles in a 12 M$_{\odot}$ $\beta$ Cep MS model (top panels) and a 4 M$_{\odot}$ SPB MS model (bottom panels) with the SMC chemical mixture and metallicity. The non-modified opacity profiles are represented by black solid lines. The profiles increased by 10 and 50 \% are drawn with long-dashed and short-dashed lines, respectively. They coincide with the normal opacity profile where no change is made, that is in central layers ($\log T$~$>$~6 ) and towards surface ($\log T$~$<$~4.5). Opacity increases in the contributions of Fe, Ni, both Cr and Mn are presented from left to right.}
\label{fig1}
\end{figure}

In Fig. \ref{exampleofprofile}, we show the normalized opacity profile (where $\kappa$ is the Rosseland mean opacity) in a 12~M$_{\odot}$ $\beta$~Cep model and the contributions of each iron group element: in the opacity bump around $\log T$~=~5.3, Fe contributes to $\sim$65~\%, Ni to $\sim$25\% and Cr and Mn to less than 15\%. Each contribution can be approximated by a Gaussian function centred at a value $\log T_i$, of width $\sigma_i$ (see Table~\ref{param}). The changes in opacity are introduced by multiplying the original profile by a Gaussian with the same profile and an amplitude increased by $A\%$ (Table~\ref{param}). Let us note that a 50\% rise in the Ni opacity corresponds to an increase of 50\% of the total opacity at the temperature ($\log T_{\rm Ni}$) at which Ni opacity has its maximum value. 
The Ni contribution alone is increased by more than a factor 2.5 since its contribution to the total opacity in that region is $\sim$~25\%.

In Fig. \ref{fig1}, we illustrate, for given structures ($\log T$, $\log \rho$, $X$, $Z$) of respectively a $\beta$ Cep model (top panels) and an SPB model (bottom panels), the effects on the opacity profile of our \emph{ad hoc} opacity modifications. Cr and Mn are treated together since their contributions are small.  As reported in Table~\ref{param}, we consequently enhance the Mn opacity by one third of the increase of the Cr opacity since the contributions of Mn and Cr to the iron group bump follows approximatively this ratio (see Fig. \ref{exampleofprofile}). Increases are always applied at the same temperatures, whatever the other quantities, on which the opacity explicitly depends, are. At a given temperature, the density ($\rho$) is not the same in an SPB as in a $\beta$ Cep structure. Consequently, the maximum of the opacity increase in SPB models is shifted to slightly higher temperatures than in $\beta$ Cep models.

\begin{figure} 
\begin{center}
\includegraphics[scale=0.35]{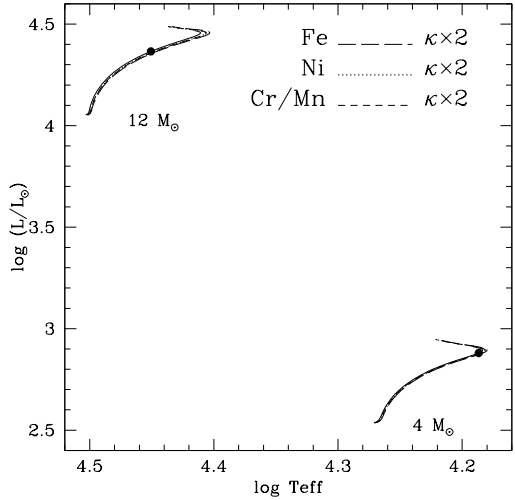} 
\end{center}
\caption{Solid lines represent evolutionary tracks from ZAMS to TAMS of 4 and 12 M$_{\odot}$ models with a normal opacity. Long-dashed, dashed and short-dashed lines show evolutionary tracks of models with increases of 100\% in the Fe, Ni, and Cr and Mn opacity, respectively. Filled circles indicate the model at $X_{\rm c}$~=~0.06 and 0.20 for the 4 and 12 M$_{\odot}$ models, respectively.}
\label{fighr} 
\end{figure}

With these modified opacities, we have computed new 4 and 12 M$_{\odot}$ models, representative of SPB and $\beta$ Cep stars, respectively, with $Z=0.00236$ and physical inputs as detailed in Sect. \ref{mssa}. This induces changes in the density and in the temperature gradient ($\nabla T$), since a modification of the opacity directly affects $\nabla T$. As a higher opacity leads to expansion of the involved stellar layers, $\rho$ decreases by 35-45\% according to the element considered and the type of star, while $\nabla T$ is affected by changes of 10-15\%. Yet, as illustrated in Fig. \ref{fighr}, effects in the Hertzsprung-Russell (HR) diagram are negligible since these changes take place only over a small fraction of the stellar radius. For a given evolutionary state, the stellar radius is increased by 1 to 2.5\% either in $\beta$ Cep or SPB models. Whilst these changes do not affect the global properties of the star, they could however have some miscellaneous effect on mode excitation. 

\subsection{Stability analysis}

\begin{figure*} 
\begin{center}
\includegraphics[scale=0.75]{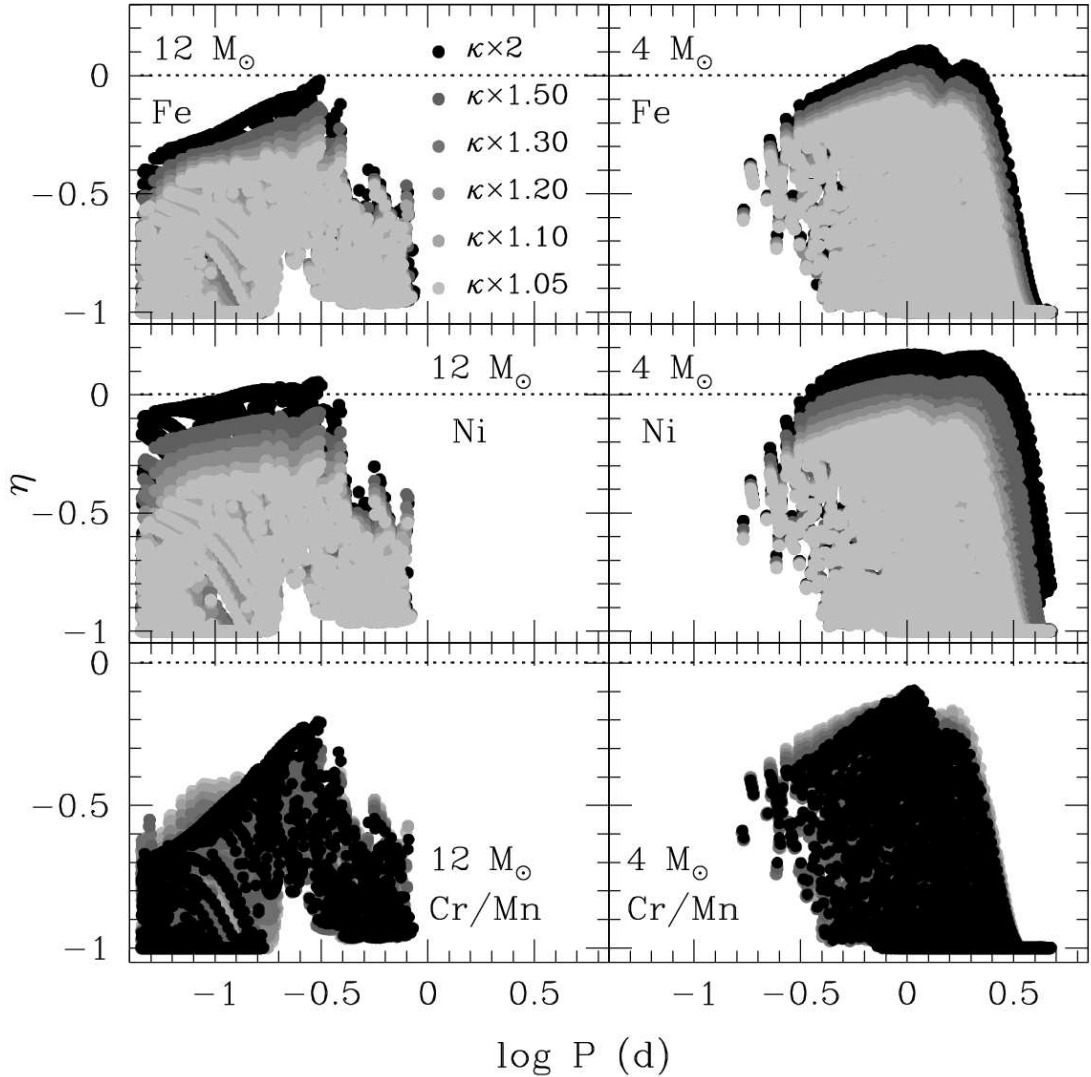} 
\end{center}
\caption{The $\eta$ parameter of modes as a function of their periods, for typical $\beta$ Cep (resp. SPB) models of the SMC stars, are shown in the left (resp. right) panel. Different levels of grey indicate different factors of opacity increase. The modification in the Fe, Ni, and combined Cr and Mn opacities are shown in the panels from top to bottom, respectively. 
} 
\label{fig2} 
\end{figure*}

We have checked the stability of these new models. The results are shown in Fig. \ref{fig2}. $\beta$ Cep modes are only excited if the opacity at the Ni maximum is increased by more than 50\% (middle left panel). Changes at the temperatures of the Fe, and Cr and Mn maxima (top and bottom left panels, respectively) do not lead to $\beta$~Cep mode excitation, even for the 100\% increase. Surprisingly, the more enhanced the opacity in Cr and Mn is, the more stable the $\beta$ Cep modes are, as shown by the decrease in $\eta$. SPB modes are excited for an increase $>$~30~\% both in the Fe and Ni cases (top and middle right panels). We note that a 30\% increase is very close to exciting SPB modes in the Ni case. No excitation of SPB modes is obtained in the Cr and Mn case, where the highest increases slightly strengthen the stability of SPB modes (bottom right panel).

The $\eta$ parameter gives information on the global stability of a mode but it does not indicate which stellar layers contribute to the driving or damping of the mode. We thus introduce local quantities that describe more precisely the $\kappa$ mechanism, according to the formalism developed in \citet{unno}, \citet{pami99} and \citet{dupret}. In particular, a stellar layer contributes to mode driving (resp. damping) if it produces a positive (resp. negative) work, i.e., $-{\rm d}W/{\rm d}m>0$ (resp. $<0$). The latter quantity can be expressed as a function of the luminosity perturbation $\delta L$:
\begin{equation}
-\frac{{\rm d}W}{{\rm d}m}\propto\Re\left\{\frac{\delta T^*}{T} \frac{d}{{\rm d}m}\frac{\delta L}{L}\right\},
\label{dw}
\end{equation}
where $\Re$ is the real part of the expression in brackets. We normalize $W$ such that its value at the surface is equal to $\Im\{\omega\}$, the imaginary part of the complex frequency of the mode.
This equation is strictly valid only for radial modes, but remains a good approximation for non-radial modes of low angular degree, $\ell$. Indeed, corrective terms are generally $\ll d/{\rm d}m (\delta L/L)$, due to the strong stratification in the radial direction. Assuming the diffusion approximation, we can further develop the right-side of Eq. (\ref{dw}) as:
\begin{eqnarray}
\Re\left\{\frac{\delta T^*}{T} \frac{d}{{\rm d}m}\frac{\delta L}{L}\right\}=\Re \left\{ \frac{\delta T^*}{T} \right. \frac{d}{{\rm d}m}&&\left[\frac{L_R}{L}\right. 
\left(4 \frac{\delta T}{T}\right.-\frac{\delta \kappa}{\kappa}+\frac{{\rm d} \delta T/T}{{\rm d} \ln T} \nonumber \\ 
&&+4\frac{\xi_r}{r}+\ell(\ell+1)\left.\left.\left.\frac{\xi_h}{r} \right)\ \right]\ \right\},
\label{diff}
\end{eqnarray}
where $L_R$ is the radiative luminosity. The term $4\xi_r/r+\ell(\ell+1)\xi_h/r $, where $\xi_r$ and $\xi_h$ represent, respectively, the radial and horizontal displacements, is the geometric variation due to the oscillation. For radial modes, $\xi_h=0$ and the term is thus sometimes called r (for radius) effect \citep{baker}. It is one to several orders of magnitude smaller than the other terms contributing to $\delta L/ L$ and does not play any significant role. The term ${\rm d} (\delta T/T)/{\rm d} \ln T$ (hereafter the temperature gradient term) generally contributes to damping of the oscillations. Driving due to the $4\ \delta T/T$ term is sometimes called the $\gamma$ mechanism \citep[see][]{coxcc}, hence we will refer to it as the $\gamma$ term. Finally, the term $-\delta \kappa/\kappa$ (hereafter the opacity term), with $\delta \kappa$ being the opacity perturbation, is the largest of these terms and the main contributor to driving in the presence of an opacity bump, giving its name to the $\kappa$ mechanism. 

The presence of an opacity bump is not sufficient for the $\kappa$ mechanism to be efficient; there are two more necessary conditions. Firstly, the opacity bump must be located in stellar layers known as the transition region, i.e. where the thermal relaxation timescale, $\tau_{{\rm th}}$, is of the same order as the period of the modes, $\Pi$. Moreover, in that region, the eigenfunctions, in particular the pressure perturbation $\delta P/P$, must have a large enough amplitude and not vary too rapidly. Given that the value of $\tau_{{\rm th}}$ decreases monotonically from centre to surface in B stars, the periods of modes that are efficiently driven increase with the depth at which the opacity bump is located, which depends on $T_{{\rm eff}}$. This is why the frequency domain as well as the type of modes are clearly distinct between SPB and $\beta$ Cep stars. Moreover, differences in the opacity profile for a given stellar model can thus alter the excitation of pulsations and the associated frequency range.

To analyse the effects of our {\it ad hoc} opacity modifications on the different terms of Eq.~\ref{diff}, we have chosen a 12~$M_\odot$ model with a central hydrogen mass fraction $X_{\rm c}=0.20$ (as a representative $\beta$~Cep star), and a 4~$M_\odot$ model before the second gravitational contraction, with $X_{\rm c}=0.06$ (as a representative SPB). The location of these two models are indicated by filled circles in Fig.~\ref{fighr}. In the next three subsections, we describe how the terms in Eq.~\ref{diff} (except for the r-term, which is negligible) react to modifications of opacity in the Fe, Ni, and Cr-Mn regions, respectively.

\subsubsection{The Fe case}

\begin{figure} 
\begin{center}
\includegraphics[scale=0.40]{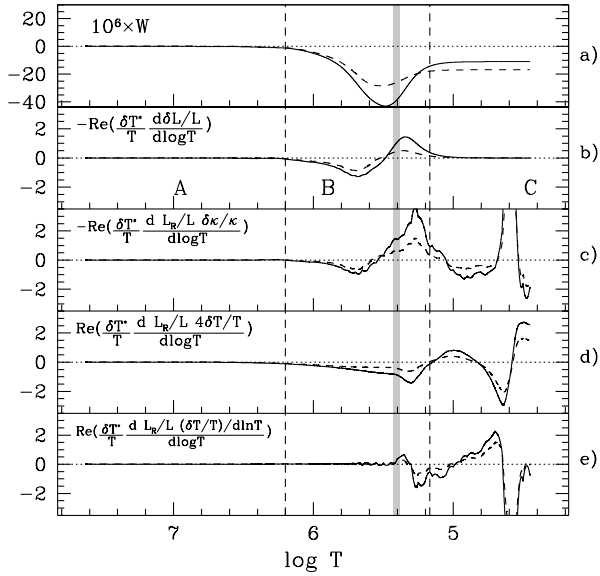} 
\end{center}
\caption{From top to bottom are respectively represented the work integral (multiplied by 10$^6$), the derivative of the work (i.e. the work developed by each stellar layer) as defined in Eq. (\ref{dw}), and the derivatives of first three terms of Eq. (\ref{diff}). The four latter quantities are multiplied by 10$^4$ for a better visibility. The long dashed (resp. continuous) line represents the $n$~=~1, $\ell$~=~0 mode of the 12 M$_{\odot}$ SMC $\beta$ Cep model computed with a normal opacity (resp. with a 100\% opacity increase at the Fe peak) and $X_{\rm c}$~=~0.204. The grey shaded area represents the transition region.
} 
\label{fig3} 
\end{figure}

In Fig.~\ref{fig3}, we show the variation inside the star of the work integral (panel a) for the fundamental radial mode of the $\beta$~Cep model.
In each panel, there are two different curves; dashed lines correspond to computations with a normal opacity, while solid ones correspond to the model with an opacity increased by 100\% in the Fe domain. The derivative of the work integral ($-{\rm d}W/{\rm d}m$) is shown in panel b, and the different terms in the right-hand side of Eq.~\ref{diff} are represented in panels c to e. 

In region A, no work is performed since we are in the quasi-adiabatic part of the star where $\delta S$~$\simeq$~0, with $\delta S$ being the entropy perturbation. On the other hand, region C has a very low heat capacity and $ {\rm d}/{\rm dm}\ (\delta L/L) \simeq 0$. There, driving and damping contributions balance to keep $\delta L/L$ constant. Finally in the stellar layers of region B, energetic processes that determine whether oscillations are efficiently driven or not, take place.

It is known \citep[see e.g.][]{unno} that the presence of an opacity bump due to an ionisation zone induces two tendencies in the derivative of the opacity. In the inner part of the ionisation zone, the derivative grows and contributes to driving the mode, as can be seen in region B (panel c) both in the normal and raised opacity cases. In the outer part of the ionisation zone, at lower temperatures, the opacity derivative decreases. This causes the opacity term to damp the oscillation in the region  4.7~$\lesssim$~$\log T$~$\lesssim$~5 (see panel c). As indicated by the negative value of $W$ at the surface (panel a), the studied mode is not excited, either in the standard 12~$M_\odot$ model or in the one with largely increased opacity. The reason can easily be understood: increasing the opacity around the Fe-maximum results in a symmetric modification of the total opacity profile, and therefore driving and damping contributions are enhanced in a similar way.
On the other hand, the moduli of the $\gamma$ and temperature gradient term contributions are enhanced in region B, leading to a stronger damping (panels d and e). As a consequence, the mode is stable even if the driving due to the $\kappa$ term increases.
Actually no low-order p or g mode (of degree $\ell=0$ to 2) is excited even when the opacity around the Fe peak is increased by a factor of 2.

\begin{table}
\centering
\caption{High-order g modes excited in the SPB model with the Fe or Ni opacity increased by 50\%. The minimal and maximal values of the period and of the $\eta$ parameter found for these modes are indicated, rounded to 2 and 4 decimal places, respectively.}
\begin{tabular}{cccc} \\\hline
Element  & $\ell$ &  P$_{{\rm min}}$--P$_{{\rm max}}$ (d) & $\eta_{\ {\rm min}}$--$\eta_{\ {\rm max}}$ \\
\hline

Fe &  2  & 1.00 -- 1.20 & 0.0027 -- 0.0221 \\    
Ni &  1  & 1.63 -- 2.47 & 0.0012 -- 0.0504\\    
   &  2  & 1.00 -- 1.49 & 0.0022 -- 0.0539\\   
  
\hline 
\end{tabular}

\label{etavalues}
\end{table}

In the SPB model with an opacity increased by 50\%, 6 high-order ($\ell=2$) g modes are excited.  Their ranges of periods and growth rates are given in Table \ref{etavalues}. The contribution of the opacity term is sufficiently enhanced to drive these modes efficiently, whereas the contributions of the other terms do not show significant differences between normal and increased opacities and do not influence the global stability of the mode.

\subsubsection{The Ni case}

Nickel is of special interest. This element is typically 30 times less abundant than Fe (see Table \ref{tab_abundances_smc}), but its contribution to the iron-group opacity bump represents about half the Fe one. Since its contribution to the opacity peak is centred at higher temperatures, it could have a relevant role in the location of the blue border of the instability strips as has already been demonstrated in sdBs \citep{jeffrey}.

\begin{figure} 
\begin{center}
\includegraphics[scale=0.40]{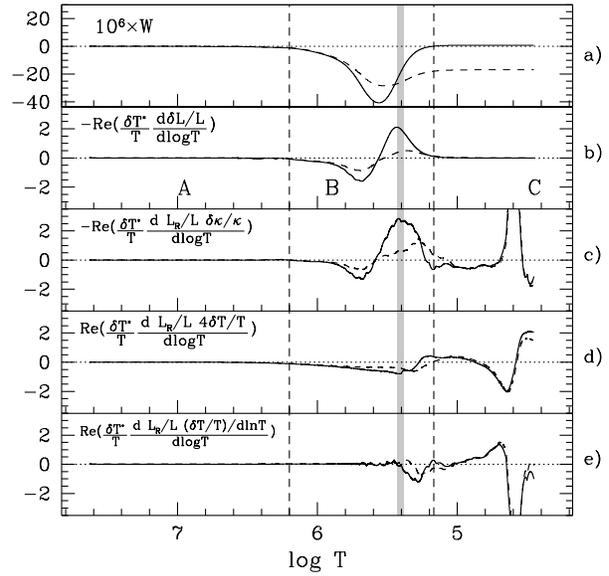} 
\end{center}
\caption{Same as Fig. \ref{fig3} but now considering a model with an opacity increased by 100\% at the Ni peak.
} 
\label{fig4} 
\end{figure}

In the $\beta$ Cep model with a Ni opacity increased by 100\%, we find the radial fundamental (n=1;$\ell$=0) and low-order (n=-1;$\ell$=1) and (n=-1;$\ell$=2) g modes to be excited. As the change is located at hotter temperatures, the driving and damping due to the $\kappa$ term are no longer affected in a symmetrical way, as was the case for Fe. In particular, raising the Ni opacity highly strengthens the driving in region B, as can be seen in panel c of Fig. \ref{fig4}. The moduli of the $\gamma$ and temperature gradient term contributions are not amplified in regards to normal opacity conditions (region B, panels d and e). Therefore the effect of the ionisation zone on pulsations is biased towards mode driving when the Ni opacity is increased, and hence the $\beta$ Cep fundamental mode is unstable ($W>0$ in panel a).

In the SPB model, increasing the opacity by 50\% leads to 15 dipole and 16 quadrupole excited high-order g modes, with periods from 1 to 2.47 d  (growth rates are indicated in Table \ref{etavalues}). The behaviour of an excited mode when the Ni opacity is increased is similar to the Fe case. The driving from the opacity term is increased so that the mode is excited while other terms are almost unaffected.

\begin{figure} 
\begin{center}
\includegraphics[scale=0.30]{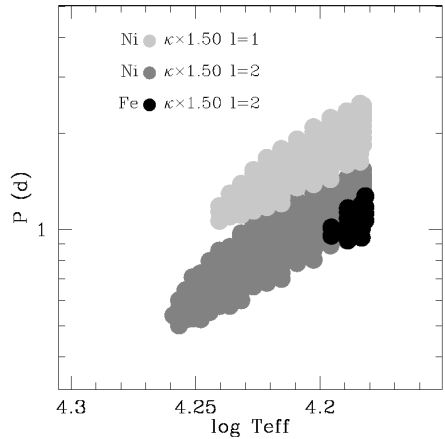} 
\end{center}
\caption{Period as a function of the effective temperature for the $\ell$=1 and $\ell$=2 excited high-order g modes in the 4 M$_{\odot}$ MS SPB models with the Ni and Fe opacities increased by 50\% (represented in light/dark grey and black filled circles, respectively). The y axis is on a logarithmic scale.}
\label{patatoide1} 
\end{figure}

However, while increasing the opacity by 50\% at the Fe peak led only to 6 quadrupole excited modes, the same enhancement in the Ni region implies the excitation of 31 SPB modes (dipole and quadrupole ones).  Furthermore, the instability domain is also affected. To illustrate this, we represent in Fig. \ref{patatoide1} the high-order g modes found to be unstable during the MS evolution of the SPB model with the Fe and the Ni opacity raised by 50\%, respectively. Firstly, the period domain is strongly increased  between the Fe and Ni cases, in particular because $\ell$=1 modes are never excited in the former case. Secondly, in the Fe case, unstable modes begin to be present at $X_{\rm c}=0.15$ ($T_{{\rm eff}}~\sim~15,900$ K) while this limit is at $X_{\rm c}=0.55$ ($T_{{\rm eff}}~\sim~18,000$ K) in the Ni case, shifting to hotter temperatures the blue border of the instability strip of SPBs.

\begin{figure} 
\begin{center}
\includegraphics[scale=0.30]{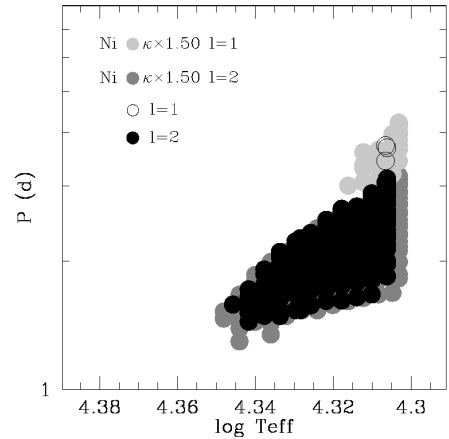} 
\end{center}
\caption{Period as a function of the effective temperature for excited high-order g modes in hybrid 10 M$_{\odot}$ MS models at Galactic metallicity ($Z$=0.014). The light and dark grey filled circles represent the modes of degree $\ell$=1 and $\ell$=2, respectively, for the stellar models computed with a Ni opacity increased by 50\%. The black open and filled circles represent the modes of degree $\ell$=1 and $\ell$=2, respectively, for the stellar models computed with a normal opacity. The y axis is on a logarithmic scale.} 
\label{patatoide3} 
\end{figure}

 As mentioned in Sect. \ref{intro}, current theoretical models are not able to explain excited low frequency modes in Galactic B-type pulsators, such as those found in 12~Lac or $\nu$~ Eri \citep{dziembo04}. The above described results motivate us to study the effect on the stability properties of a 10 M$_{\odot}$ hybrid model with typical Galactic metallicity ($Z$~=~0.014) of an enhancement of opacity by 50\% at the Ni peak. The periods of the unstable high-order g (SPB) modes are represented in Fig. \ref{patatoide3} by black open ($\ell$=1) or filled ($\ell$=2) circles for the normal case and light ($\ell$=1) or dark ($\ell$=2) grey filled circles for the enhanced Ni case. The $\ell$=2 modes present a similar instability domain. However the $\ell$=1 domain is significantly increased in temperature and in period when the Ni opacity is raised. Hence there is a larger number of unstable modes with high period values, which are usually the ones not reproduced when modelling Galactic hybrids. Finally, the instability domain of $\beta$ Cep modes (which is not represented) remains unchanged from normal opacity conditions.

\subsubsection{The Cr and Mn case}

Increasing the Cr and Mn opacities does not allow either the $\beta$ Cep or the SPB modes to be driven. More surprisingly, growth rates are weaker when the enhancements are larger (bottom panels of Fig. \ref{fig2}).

\begin{figure} 
\begin{center}
\includegraphics[scale=0.40]{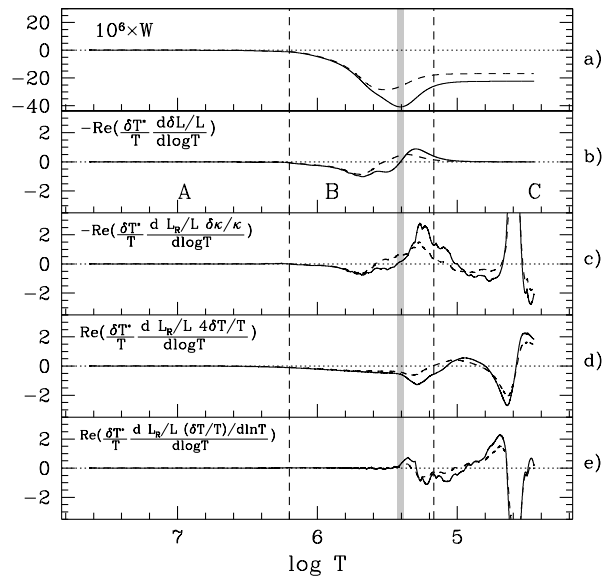} 
\end{center}
\caption{Same as Fig. \ref{fig3} but now considering a model with an opacity increased by 100\% at the Cr peak and 100/3 \% at the Mn peak.
} 
\label{fig5} 
\end{figure}

We can understand this result from the opacity term (panel c of Fig. \ref{fig5}) for the $\beta$ Cep fundamental mode. The rise in opacity moves the driving component of the opacity term towards cooler temperatures outside the transition region of the fundamental mode. Since $\delta L/L$ is constant in these low heat capacity regions, the driving is balanced by the $\gamma$ and temperature gradient terms. In such layers, every energy gain is directly lost as heat, inhibiting the conversion of the pulsation into mechanical energy.

In the SPB model, with the increase of the opacity, the modulus of the opacity term contribution vanishes in layers where it was driving in the normal case. Hence the growth rate of the SPB modes is also decreased as the Cr and Mn opacities are enhanced.

\section{Discussion}

We have further investigated the consequences of increasing the Fe or Ni opacities in models where metallicities of $Z=0.003$, 0.004 and 0.005 are now considered. In the Fe case, no $\beta$ Cep modes are excited for $Z=0.003$ even for an increase of 100\%. However, $\beta$ Cep modes are excited at $Z=0.004$ and $Z=0.005$ when 100\% and 50\% enhancements are applied, respectively. In the Ni case, for $Z=0.003$, 0.004 and 0.005 we need increases of at least 100, 100 and 50 \%, respectively, to excite $\beta$ Cep pulsations. Considering SPB models, we obtain excited high-order g modes at $Z=0.003$ for opacity increases of 20\% and 30\% in Ni and Fe, respectively. At $Z=0.004$ an increase of 5\% yields unstable modes both in the Ni and Fe cases, as expected from results with the SMC mixture analysis in Sect. \ref{mssa} (see Fig. \ref{bandinst}).

We note that a determination of abundances for several Magellanic B-type stars, including pulsator candidates, would allow us to assess if there is a clear deviation from the mean abundances --in particular the ones of iron group elements-- we have adopted in this work.
Though it would be valuable to establish the status of SPB and $\beta$ Cep candidates in the MCs, our work also presents interesting perspectives for Galactic hybrid B-type pulsators.  Additional calculations for a Galactic model of a hybrid $\beta$ Cep/SPB star showed that, while the $\beta$ Cep domain instability domain remains unchanged, the SPB domain is broadened when Ni opacity is increased.

\section{Conclusion}
\label{lastpage}

In this paper we search for an explanation for the presence of main sequence B-type pulsators in low-metallicity environments such as the Magellanic Clouds. Our investigation follows two approaches. First, we show that representative metal mixtures for B stars are not able to explain the presence of SPBs and especially of $\beta$ Cephei stars in the more metal poor galaxy, i.e. in the Small Magellanic Cloud.


In our second approach, we examine the effects of increasing the opacities of Fe, Ni, Cr and Mn in stellar models representative of the Small Magellanic Cloud. It is not possible to excite both SPB and $\beta$ Cep modes with an enhancement of the opacity at the Fe peak, even for the largest increase. However, an increase of more than 50~\% of the opacity profile at the Ni peak leads to the excitation of both SPB and $\beta$ Cep modes. Finally, simultaneously raising the Cr and Mn opacities does not increase the probability of driving modes, instead it reinforces the stability of main sequence B stars.

We also show that a modification of the opacity in the Ni region moves the blue border of the SPBs instability strip to hotter temperatures and extends the frequency domain of pulsations, as expected from previous studies \citep{pami99,miglio}.

Neither OPAL nor OP stellar opacities are able to reproduce the lowest frequencies (i.e. the highest periods) observed in Galactic hybrid B-type pulsators. A similar outcome in the modelling of sdB stars already led \citet{jeffrey} to suggest an increase of the Ni contribution to opacity. Moreover, Ni (as well as Cr and Mn) atomic data used by OP to compute stellar opacities are not completely obtained \emph{ab initio}. Instead, extrapolations from Fe results are adopted \citep{badnell}. These uncertainties in the iron-group opacities are at the origin of a multi-disciplinary collaboration (see Gilles et al. 2011, for preliminary results)\nocite{Gilles112}, with the aim of comparing theoretical codes for opacity computations and carrying out experiments to measure opacities in conditions equivalent to those of stellar plasmas.

Moreover, in the present study we find that an enhancement of the opacity in the region where Ni contributes could explain the presence of B-type pulsators in the SMC.   Raising the Ni opacity also produces a larger period domain of excited modes in models of Galactic hybrid B-type pulsators, extending it in particular to higher periods (i.e. to lower frequencies) as required to explain observational results. Already in the past, Simon (1982) \nocite{simon} made a plea for a reexamination of heavy element opacities. This was verified in the following decade when opacity due to the iron group increased by a factor of 3 for B star conditions. {\it We hence conclude that B-type pulsators in the Magellanic Clouds and in the Milky Way suggest that current opacity data underestimate stellar opacity due to Ni by a factor $\lesssim$2. Therefore, we call for a reexamination of Ni opacity calculations.}


\section*{Acknowledgments}

M-P. Bouabid is aknowledged for his stimulating discussions on this work. D. Reese is warmly thanked for his revision of the English language. S.S. works under PhD grant from F.R.I.A., Fonds pour la formation \`{a} la Recherche dans l'Industrie et l'Agriculture, Belgium. J.M. and T.M. acknowledge financial support from Belspo for contracts COROT and GAIA-DPAC, respectively.

\bibliographystyle{mn2e}
\bibliography{biblio}

\newcommand{\noopsort}[1]{}
\begin{thebibliography}{}

\bibitem[\protect\citeauthoryear{{Aerts}, {Thoul}, {Daszy{\'n}ska},
  {Scuflaire}, {Waelkens}, {Dupret}, {Niemczura} \& {Noels}}{{Aerts}
  et~al.}{2003}]{Aerts03}
{Aerts} C.,  {Thoul} A.,  {Daszy{\'n}ska} J.,  {Scuflaire} R.,  {Waelkens} C.,
  {Dupret} M.~A.,  {Niemczura} E.,    {Noels} A.,  2003, Science, 300, 1926

\bibitem[\protect\citeauthoryear{{Andrievsky}, {Kovtyukh}, {Korotin}, {Spite}
  \& {Spite}}{{Andrievsky} et~al.}{2001}]{andrievsky01}
{Andrievsky} S.~M.,  {Kovtyukh} V.~V.,  {Korotin} S.~A.,  {Spite} M.,
  {Spite} F.,  2001, \aap, 367, 605

\bibitem[\protect\citeauthoryear{{Angulo C. et al.,}}{{Angulo C. et
  al.,}}{1999}]{angulo}
{Angulo C. et al.,} 1999, Nuclear Physics A, 656, 3

\bibitem[\protect\citeauthoryear{{Asplund}, {Grevesse} \& {Sauval}}{{Asplund}
  et~al.}{2005}]{asplund}
{Asplund} M.,  {Grevesse} N.,    {Sauval} A.~J.,  2005, in {T.~G.~Barnes III \&
  F.~N.~Bash} ed., Cosmic Abundances as Records of Stellar Evolution and
  Nucleosynthesis Vol.~336 of Astronomical Society of the Pacific Conference
  Series, {The Solar Chemical Composition}.
p.~25

\bibitem[\protect\citeauthoryear{{Asplund}, {Grevesse}, {Sauval} \&
  {Scott}}{{Asplund} et~al.}{2009}]{AGS09}
{Asplund} M.,  {Grevesse} N.,  {Sauval} A.~J.,    {Scott} P.,  2009, \araa, 47,
  481

\bibitem[\protect\citeauthoryear{{Ausseloos}, {Scuflaire}, {Thoul} \&
  {Aerts}}{{Ausseloos} et~al.}{2004}]{Ausseloos04}
{Ausseloos} M.,  {Scuflaire} R.,  {Thoul} A.,    {Aerts} C.,  2004, \mnras,
  355, 352

\bibitem[\protect\citeauthoryear{{Badnell}, {Bautista}, {Butler}, {Delahaye},
  {Mendoza}, {Palmeri}, {Zeippen} \& {Seaton}}{{Badnell}
  et~al.}{2005}]{badnell}
{Badnell} N.~R.,  {Bautista} M.~A.,  {Butler} K.,  {Delahaye} F.,  {Mendoza}
  C.,  {Palmeri} P.,  {Zeippen} C.~J.,    {Seaton} M.~J.,  2005, \mnras, 360,
  458

\bibitem[\protect\citeauthoryear{{Bahcall}, {Basu}, {Pinsonneault} \&
  {Serenelli}}{{Bahcall} et~al.}{2005}]{bahcall}
{Bahcall} J.~N.,  {Basu} S.,  {Pinsonneault} M.,    {Serenelli} A.~M.,  2005,
  \apj, 618, 1049

\bibitem[\protect\citeauthoryear{{Baker}}{{Baker}}{1966}]{baker}
{Baker} N.,  1966, in {R.~F.~Stein \& A.~G.~W.~Cameron} ed., Stellar Evolution
  {Simplified Models for Cepheid Instability}.
p.~333

\bibitem[\protect\citeauthoryear{{B{\"o}hm-Vitense}}{{B{\"o}hm-Vitense}}{1958}%
]{bohm}
{B{\"o}hm-Vitense} E.,  1958, \zap, 46, 108

\bibitem[\protect\citeauthoryear{{Briquet}, {Morel}, {Thoul}, {Scuflaire},
  {Miglio}, {Montalb{\'a}n}, {Dupret} \& {Aerts}}{{Briquet}
  et~al.}{2007}]{Briquet07}
{Briquet} M.,  {Morel} T.,  {Thoul} A.,  {Scuflaire} R.,  {Miglio} A.,
  {Montalb{\'a}n} J.,  {Dupret} M.-A.,    {Aerts} C.,  2007, \mnras, 381, 1482

\bibitem[\protect\citeauthoryear{{Buchler}}{{Buchler}}{2008}]{buchler}
{Buchler} J.~R.,  2008, \apj, 680, 1412

\bibitem[\protect\citeauthoryear{{Carrera}, {Gallart}, {Aparicio}, {Costa},
  {M{\'e}ndez} \& {No{\"e}l}}{{Carrera} et~al.}{2008}]{carrera}
{Carrera} R.,  {Gallart} C.,  {Aparicio} A.,  {Costa} E.,  {M{\'e}ndez} R.~A.,
    {No{\"e}l} N.~E.~D.,  2008, \aj, 136, 1039

\bibitem[\protect\citeauthoryear{{Cox}, {Morgan}, {Rogers} \& {Iglesias}}{{Cox}
  et~al.}{1992}]{cox}
{Cox} A.~N.,  {Morgan} S.~M.,  {Rogers} F.~J.,    {Iglesias} C.~A.,  1992,
  \apj, 393, 272

\bibitem[\protect\citeauthoryear{{Cox}, {Cox}, {Olsen}, {King} \&
  {Eilers}}{{Cox} et~al.}{1966}]{coxcc}
{Cox} J.~P.,  {Cox} A.~N.,  {Olsen} K.~H.,  {King} D.~S.,    {Eilers} D.~D.,
  1966, \apj, 144, 1038

\bibitem[\protect\citeauthoryear{{Daszy{\'n}ska-Daszkiewicz} \&
  {Walczak}}{{Daszy{\'n}ska-Daszkiewicz} \& {Walczak}}{2009}]{jagoda09}
{Daszy{\'n}ska-Daszkiewicz} J.,  {Walczak} P.,  2009, \mnras, 398, 1961

\bibitem[\protect\citeauthoryear{{Daszy{\'n}ska-Daszkiewicz} \&
  {Walczak}}{{Daszy{\'n}ska-Daszkiewicz} \& {Walczak}}{2010}]{jagoda10}
{Daszy{\'n}ska-Daszkiewicz} J.,  {Walczak} P.,  2010, \mnras, 403, 496

\bibitem[\protect\citeauthoryear{{De Propris}, {Rich}, {Mallery} \&
  {Howard}}{{De Propris} et~al.}{2010}]{depropis}
{De Propris} R.,  {Rich} R.~M.,  {Mallery} R.~C.,    {Howard} C.~D.,  2010,
  \apjl, 714, L249

\bibitem[\protect\citeauthoryear{{Desmet}, {Briquet}, {Thoul}, {Zima}, {De
  Cat}, {Handler}, {Ilyin}, {Kambe}, {Krzesinski}, {Lehmann}, {Masuda},
  {Mathias}, {Mkrtichian}, {Telting}, {Uytterhoeven}, {Yang} \&
  {Aerts}}{{Desmet} et~al.}{2009}]{Desmet09}
{Desmet} M.,  {Briquet} M.,  {Thoul} A.,  {Zima} W.,  {De Cat} P.,  {Handler}
  G.,  {Ilyin} I.,  {Kambe} E.,  {Krzesinski} J.,  {Lehmann} H.,  {Masuda} S.,
  {Mathias} P.,  {Mkrtichian} D.~E.,  {Telting} J.,  {Uytterhoeven} K.,  {Yang}
  S.~L.~S.,    {Aerts} C.,  2009, \mnras, 396, 1460

\bibitem[\protect\citeauthoryear{{Diago}, {Guti{\'e}rrez-Soto}, {Fabregat} \&
  {Martayan}}{{Diago} et~al.}{2008}]{diago}
{Diago} P.~D.,  {Guti{\'e}rrez-Soto} J.,  {Fabregat} J.,    {Martayan} C.,
  2008, \aap, 480, 179

\bibitem[\protect\citeauthoryear{{Dupret}}{{Dupret}}{2001}]{dupret}
{Dupret} M.~A.,  2001, \aap, 366, 166

\bibitem[\protect\citeauthoryear{{Dupret}, {Thoul}, {Scuflaire},
  {Daszy{\'n}ska-Daszkiewicz}, {Aerts}, {Bourge}, {Waelkens} \&
  {Noels}}{{Dupret} et~al.}{2004}]{Dupret04}
{Dupret} M.-A.,  {Thoul} A.,  {Scuflaire} R.,  {Daszy{\'n}ska-Daszkiewicz} J.,
  {Aerts} C.,  {Bourge} P.-O.,  {Waelkens} C.,    {Noels} A.,  2004, \aap, 415,
  251

\bibitem[\protect\citeauthoryear{{Dziembowski} \& {Jerzykiewicz}}{{Dziembowski}
  \& {Jerzykiewicz}}{1996}]{Jerki96}
{Dziembowski} W.~A.,  {Jerzykiewicz} M.,  1996, \aap, 306, 436

\bibitem[\protect\citeauthoryear{{Dziembowski}, {Moskalik} \&
  {Pamyatnykh}}{{Dziembowski} et~al.}{1993}]{dziembowski93}
{Dziembowski} W.~A.,  {Moskalik} P.,    {Pamyatnykh} A.~A.,  1993, \mnras, 265,
  588

\bibitem[\protect\citeauthoryear{{Dziembowski} \& {Pamyatnykh}}{{Dziembowski}
  \& {Pamyatnykh}}{2008}]{dziembowski4}
{Dziembowski} W.~A.,  {Pamyatnykh} A.~A.,  2008, \mnras, 385, 2061

\bibitem[\protect\citeauthoryear{{Esteban}, {Peimbert}, {Torres-Peimbert} \&
  {Escalante}}{{Esteban} et~al.}{1998}]{esteban98}
{Esteban} C.,  {Peimbert} M.,  {Torres-Peimbert} S.,    {Escalante} V.,  1998,
  \mnras, 295, 401

\bibitem[\protect\citeauthoryear{{Formicola A. et al.,}}{{Formicola A. et
  al.,}}{2004}]{formicola}
{Formicola A. et al.,} 2004, Physics Letters B, 591, 61

\bibitem[\protect\citeauthoryear{{Garnett}}{{Garnett}}{1999}]{garnett99}
{Garnett} D.~R.,  1999, in {Y.-H.~Chu, N.~Suntzeff, J.~Hesser, \& D.~Bohlender}
  ed., New Views of the Magellanic Clouds Vol.~190 of IAU Symposium, {Element
  Abundances in Magellanic Cloud H II Regions from Carbon to Argon}.
p.~266

\bibitem[\protect\citeauthoryear{{Gilles D. et al.,}}{{Gilles D. et
  al.,}}{2011}]{Gilles112}
{Gilles D. et al.,} 2011, High Energy Density Physics, 7, 312

\bibitem[\protect\citeauthoryear{{Gonzalez} \& {Wallerstein}}{{Gonzalez} \&
  {Wallerstein}}{1999}]{gonzalez99}
{Gonzalez} G.,  {Wallerstein} G.,  1999, \aj, 117, 2286

\bibitem[\protect\citeauthoryear{{Grevesse} \& {Noels}}{{Grevesse} \&
  {Noels}}{1993}]{gn93}
{Grevesse} N.,  {Noels} A.,  1993, in {N.~Prantzos, E.~Vangioni-Flam, \&
  M.~Casse} ed., Origin and Evolution of the Elements {Cosmic abundances of the
  elements.}.
pp 15--25

\bibitem[\protect\citeauthoryear{{Hill}}{{Hill}}{1997}]{hill97}
{Hill} V.,  1997, \aap, 324, 435

\bibitem[\protect\citeauthoryear{{Hill}}{{Hill}}{1999}]{hill99}
{Hill} V.,  1999, \aap, 345, 430

\bibitem[\protect\citeauthoryear{{Hill}, {Barbuy} \& {Spite}}{{Hill}
  et~al.}{1997}]{hilletal97}
{Hill} V.,  {Barbuy} B.,    {Spite} M.,  1997, \aap, 323, 461

\bibitem[\protect\citeauthoryear{{Hunter}, {Brott}, {Lennon}, {Langer},
  {Dufton}, {Trundle}, {Smartt}, {de Koter}, {Evans} \& {Ryans}}{{Hunter}
  et~al.}{2008}]{hunter08}
{Hunter} I.,  {Brott} I.,  {Lennon} D.~J.,  {Langer} N.,  {Dufton} P.~L.,
  {Trundle} C.,  {Smartt} S.~J.,  {de Koter} A.,  {Evans} C.~J.,    {Ryans}
  R.~S.~I.,  2008, \apjl, 676, L29

\bibitem[\protect\citeauthoryear{{Hunter}, {Dufton}, {Ryans}, {Lennon},
  {Rolleston}, {Hubeny} \& {Lanz}}{{Hunter} et~al.}{2005}]{hunter05}
{Hunter} I.,  {Dufton} P.~L.,  {Ryans} R.~S.~I.,  {Lennon} D.~J.,  {Rolleston}
  W.~R.~J.,  {Hubeny} I.,    {Lanz} T.,  2005, \aap, 436, 687

\bibitem[\protect\citeauthoryear{{Hunter}, {Dufton}, {Smartt}, {Ryans},
  {Evans}, {Lennon}, {Trundle}, {Hubeny} \& {Lanz}}{{Hunter}
  et~al.}{2007}]{hunter07}
{Hunter} I.,  {Dufton} P.~L.,  {Smartt} S.~J.,  {Ryans} R.~S.~I.,  {Evans}
  C.~J.,  {Lennon} D.~J.,  {Trundle} C.,  {Hubeny} I.,    {Lanz} T.,  2007,
  \aap, 466, 277

\bibitem[\protect\citeauthoryear{{Idiart}, {Maciel} \& {Costa}}{{Idiart}
  et~al.}{2007}]{idiart}
{Idiart} T.~P.,  {Maciel} W.~J.,    {Costa} R.~D.~D.,  2007, \aap, 472, 101

\bibitem[\protect\citeauthoryear{{Iglesias} \& {Rogers}}{{Iglesias} \&
  {Rogers}}{1996}]{rogers96}
{Iglesias} C.~A.,  {Rogers} F.~J.,  1996, \apj, 464, 943

\bibitem[\protect\citeauthoryear{{Jeffery} \& {Saio}}{{Jeffery} \&
  {Saio}}{2006}]{jeffrey}
{Jeffery} C.~S.,  {Saio} H.,  2006, \mnras, 372, L48

\bibitem[\protect\citeauthoryear{{Karoff}, {Arentoft}, {Glowienka}, {Coutures},
  {Nielsen}, {Dogan}, {Grundahl} \& {Kjeldsen}}{{Karoff} et~al.}{2008}]{karoff}
{Karoff} C.,  {Arentoft} T.,  {Glowienka} L.,  {Coutures} C.,  {Nielsen} T.~B.,
   {Dogan} G.,  {Grundahl} F.,    {Kjeldsen} H.,  2008, \mnras, 386, 1085

\bibitem[\protect\citeauthoryear{{Ko{\l}aczkowski}, {Pigulski},
  {Soszy{\'n}ski}, {Udalski}, {Kubiak}, {Szyma{\'n}ski}, {{\.Z}ebru{\'n}},
  {Pietrzy{\'n}ski}, {Wo{\'z}niak}, {Szewczyk} \&
  {Wyrzykowski}}{{Ko{\l}aczkowski} et~al.}{2006}]{Kola}
{Ko{\l}aczkowski} Z.,  {Pigulski} A.,  {Soszy{\'n}ski} I.,  {Udalski} A.,
  {Kubiak} M.,  {Szyma{\'n}ski} M.,  {{\.Z}ebru{\'n}} K.,  {Pietrzy{\'n}ski}
  G.,  {Wo{\'z}niak} P.~R.,  {Szewczyk} O.,    {Wyrzykowski} {\L}.,  2006,
  Memorie della Societa Astronomica Italiana, 77, 336

\bibitem[\protect\citeauthoryear{{Korn}, {Becker}, {Gummersbach} \&
  {Wolf}}{{Korn} et~al.}{2000}]{korn00}
{Korn} A.~J.,  {Becker} S.~R.,  {Gummersbach} C.~A.,    {Wolf} B.,  2000, \aap,
  353, 655

\bibitem[\protect\citeauthoryear{{Korn}, {Keller}, {Kaufer}, {Langer},
  {Przybilla}, {Stahl} \& {Wolf}}{{Korn} et~al.}{2002}]{korn02}
{Korn} A.~J.,  {Keller} S.~C.,  {Kaufer} A.,  {Langer} N.,  {Przybilla} N.,
  {Stahl} O.,    {Wolf} B.,  2002, \aap, 385, 143

\bibitem[\protect\citeauthoryear{{Korn}, {Nieva}, {Daflon} \& {Cunha}}{{Korn}
  et~al.}{2005}]{korn05}
{Korn} A.~J.,  {Nieva} M.~F.,  {Daflon} S.,    {Cunha} K.,  2005, \apj, 633,
  899

\bibitem[\protect\citeauthoryear{{Kurt}, {Dufour}, {Garnett}, {Skillman},
  {Mathis}, {Peimbert}, {Torres-Peimbert} \& {Ruiz}}{{Kurt}
  et~al.}{1999}]{kurt99}
{Kurt} C.~M.,  {Dufour} R.~J.,  {Garnett} D.~R.,  {Skillman} E.~D.,  {Mathis}
  J.~S.,  {Peimbert} M.,  {Torres-Peimbert} S.,    {Ruiz} M.-T.,  1999, \apj,
  518, 246

\bibitem[\protect\citeauthoryear{{Lebouteiller}, {Bernard-Salas}, {Brandl},
  {Whelan}, {Wu}, {Charmandaris}, {Devost} \& {Houck}}{{Lebouteiller}
  et~al.}{2008}]{lebouteiller08}
{Lebouteiller} V.,  {Bernard-Salas} J.,  {Brandl} B.,  {Whelan} D.~G.,  {Wu}
  Y.,  {Charmandaris} V.,  {Devost} D.,    {Houck} J.~R.,  2008, \apj, 680, 398

\bibitem[\protect\citeauthoryear{{Lennon}, {Dufton} \& {Crowley}}{{Lennon}
  et~al.}{2003}]{lennon03}
{Lennon} D.~J.,  {Dufton} P.~L.,    {Crowley} C.,  2003, \aap, 398, 455

\bibitem[\protect\citeauthoryear{Loisel, Arnault, Bastiani-Ceccotti, Blenski,
  Caillaud, Fariaut, Fölsner, Gilleron, Pain, Poirier, Reverdin, Silvert,
  Thais, Turck-Chièze \& Villette}{Loisel et~al.}{2009}]{TC3}
Loisel G.,  Arnault P.,  Bastiani-Ceccotti S.,  Blenski T.,  Caillaud T.,
  Fariaut J.,  Fölsner W.,  Gilleron F.,  Pain J.-C.,  Poirier M.,  Reverdin
  C.,  Silvert V.,  Thais F.,  Turck-Chièze S.,    Villette B.,  2009, High
  Energy Density Physics, 5, 173

\bibitem[\protect\citeauthoryear{{Luck}, {Moffett}, {Barnes} III \&
  {Gieren}}{{Luck} et~al.}{1998}]{luck98}
{Luck} R.~E.,  {Moffett} T.~J.,  {Barnes} III T.~G.,    {Gieren} W.~P.,  1998,
  \aj, 115, 605

\bibitem[\protect\citeauthoryear{{Mazumdar}, {Briquet}, {Desmet} \&
  {Aerts}}{{Mazumdar} et~al.}{2006}]{Mazumdar06}
{Mazumdar} A.,  {Briquet} M.,  {Desmet} M.,    {Aerts} C.,  2006, \aap, 459,
  589

\bibitem[\protect\citeauthoryear{{Miglio}, {Montalb{\'a}n} \&
  {Dupret}}{{Miglio} et~al.}{2007a}]{mmma}
{Miglio} A.,  {Montalb{\'a}n} J.,    {Dupret} M.-A.,  2007a, \mnras, 375, L21

\bibitem[\protect\citeauthoryear{{Miglio}, {Montalb{\'a}n} \&
  {Dupret}}{{Miglio} et~al.}{2007b}]{miglio}
{Miglio} A.,  {Montalb{\'a}n} J.,    {Dupret} M.-A.,  2007b, Communications in
  Asteroseismology, 151, 48

\bibitem[\protect\citeauthoryear{{Montalb{\'a}n} \& {Miglio}}{{Montalb{\'a}n}
  \& {Miglio}}{2008}]{montalban44tau}
{Montalb{\'a}n} J.,  {Miglio} A.,  2008, Communications in Asteroseismology,
  157, 160

\bibitem[\protect\citeauthoryear{{Montalb{\'a}n}, {Miglio} \&
  {Morel}}{{Montalb{\'a}n} et~al.}{2009}]{montalbanliege}
{Montalb{\'a}n} J.,  {Miglio} A.,    {Morel} T.,  2009, Communications in
  Asteroseismology, 158, 288

\bibitem[\protect\citeauthoryear{{Montalb{\'a}n}, {Miglio}, {Noels}, {Grevesse}
  \& {di Mauro}}{{Montalb{\'a}n} et~al.}{2004}]{montalbansun}
{Montalb{\'a}n} J.,  {Miglio} A.,  {Noels} A.,  {Grevesse} N.,    {di Mauro}
  M.~P.,  2004, in {D.~Danesy} ed., SOHO 14 Helio- and Asteroseismology:
  Towards a Golden Future Vol.~559 of ESA Special Publication, {Solar Model
  with CNO Revised Abundances}.
p.~574

\bibitem[\protect\citeauthoryear{{Moskalik} \& {Dziembowski}}{{Moskalik} \&
  {Dziembowski}}{1992}]{dziembowski92}
{Moskalik} P.,  {Dziembowski} W.~A.,  1992, \aap, 256, L5

\bibitem[\protect\citeauthoryear{{Pamyatnykh}}{{Pamyatnykh}}{1999}]{pami99}
{Pamyatnykh} A.~A.,  1999, Acta Astronomica, 49, 119

\bibitem[\protect\citeauthoryear{{Pamyatnykh}}{{Pamyatnykh}}{2007}]{pami07}
{Pamyatnykh} A.~A.,  2007, Communications in Asteroseismology, 150, 207

\bibitem[\protect\citeauthoryear{{Pamyatnykh}, {Handler} \&
  {Dziembowski}}{{Pamyatnykh} et~al.}{2004}]{dziembo04}
{Pamyatnykh} A.~A.,  {Handler} G.,    {Dziembowski} W.~A.,  2004, \mnras, 350,
  1022

\bibitem[\protect\citeauthoryear{{Parisi}, {Geisler}, {Grocholski},
  {Clari{\'a}} \& {Sarajedini}}{{Parisi} et~al.}{2010}]{parisi2}
{Parisi} M.~C.,  {Geisler} D.,  {Grocholski} A.~J.,  {Clari{\'a}} J.~J.,
  {Sarajedini} A.,  2010, \aj, 139, 1168

\bibitem[\protect\citeauthoryear{{Parisi}, {Grocholski}, {Geisler},
  {Sarajedini} \& {Clari{\'a}}}{{Parisi} et~al.}{2009}]{parisi1}
{Parisi} M.~C.,  {Grocholski} A.~J.,  {Geisler} D.,  {Sarajedini} A.,
  {Clari{\'a}} J.~J.,  2009, \aj, 138, 517

\bibitem[\protect\citeauthoryear{{Peimbert}}{{Peimbert}}{2003}]{peimbert03}
{Peimbert} A.,  2003, \apj, 584, 735

\bibitem[\protect\citeauthoryear{{Piatti}, {Sarajedini}, {Geisler}, {Clark} \&
  {Seguel}}{{Piatti} et~al.}{007a}]{piattia}
{Piatti} A.~E.,  {Sarajedini} A.,  {Geisler} D.,  {Clark} D.,    {Seguel} J.,
  2007a, \mnras, 377, 300

\bibitem[\protect\citeauthoryear{{Piatti}, {Sarajedini}, {Geisler}, {Gallart}
  \& {Wischnjewsky}}{{Piatti} et~al.}{007b}]{piattic}
{Piatti} A.~E.,  {Sarajedini} A.,  {Geisler} D.,  {Gallart} C.,
  {Wischnjewsky} M.,  2007b, \mnras, 381, L84

\bibitem[\protect\citeauthoryear{{Reyes}}{{Reyes}}{1999}]{reyes99}
{Reyes} R.~E.~C.,  1999, in {Y.-H.~Chu, N.~Suntzeff, J.~Hesser, \&
  D.~Bohlender} ed., New Views of the Magellanic Clouds Vol.~190 of IAU
  Symposium, {Chemical Abundances and Physical Parameters of H II Regions in
  the Magellanic Clouds}.
p.~282

\bibitem[\protect\citeauthoryear{{Rogers} \& {Iglesias}}{{Rogers} \&
  {Iglesias}}{1992}]{iglesias}
{Rogers} F.~J.,  {Iglesias} C.~A.,  1992, \apjs, 79, 507

\bibitem[\protect\citeauthoryear{{Rogers} \& {Nayfonov}}{{Rogers} \&
  {Nayfonov}}{2002}]{rogers}
{Rogers} F.~J.,  {Nayfonov} A.,  2002, \apj, 576, 1064

\bibitem[\protect\citeauthoryear{{Rolleston}, {Brown}, {Dufton} \&
  {Howarth}}{{Rolleston} et~al.}{1996}]{rolleston96}
{Rolleston} W.~R.~J.,  {Brown} P.~J.~F.,  {Dufton} P.~L.,    {Howarth} I.~D.,
  1996, \aap, 315, 95

\bibitem[\protect\citeauthoryear{{Rolleston}, {Trundle} \&
  {Dufton}}{{Rolleston} et~al.}{2002}]{rolleston}
{Rolleston} W.~R.~J.,  {Trundle} C.,    {Dufton} P.~L.,  2002, \aap, 396, 53

\bibitem[\protect\citeauthoryear{{Romaniello}, {Primas}, {Mottini},
  {Pedicelli}, {Lemasle}, {Bono}, {Fran{\c c}ois}, {Groenewegen} \&
  {Laney}}{{Romaniello} et~al.}{2008}]{romaniello08}
{Romaniello} M.,  {Primas} F.,  {Mottini} M.,  {Pedicelli} S.,  {Lemasle} B.,
  {Bono} G.,  {Fran{\c c}ois} P.,  {Groenewegen} M.~A.~T.,    {Laney} C.~D.,
  2008, \aap, 488, 731

\bibitem[\protect\citeauthoryear{{Russell} \& {Bessell}}{{Russell} \&
  {Bessell}}{1989}]{russel89}
{Russell} S.~C.,  {Bessell} M.~S.,  1989, \apjs, 70, 865

\bibitem[\protect\citeauthoryear{{Sarro}, {Debosscher}, {L{\'o}pez} \&
  {Aerts}}{{Sarro} et~al.}{2009}]{sarro}
{Sarro} L.~M.,  {Debosscher} J.,  {L{\'o}pez} M.,    {Aerts} C.,  2009, \aap,
  494, 739

\bibitem[\protect\citeauthoryear{{Scuflaire}, {Th{\'e}ado}, {Montalb{\'a}n},
  {Miglio}, {Bourge}, {Godart}, {Thoul} \& {Noels}}{{Scuflaire}
  et~al.}{2008}]{scuflaire}
{Scuflaire} R.,  {Th{\'e}ado} S.,  {Montalb{\'a}n} J.,  {Miglio} A.,  {Bourge}
  P.-O.,  {Godart} M.,  {Thoul} A.,    {Noels} A.,  2008, \apss, 316, 83

\bibitem[\protect\citeauthoryear{{Seaton} \& {Badnell}}{{Seaton} \&
  {Badnell}}{2004}]{seaton}
{Seaton} M.~J.,  {Badnell} N.~R.,  2004, \mnras, 354, 457

\bibitem[\protect\citeauthoryear{{Seaton}, {Yan}, {Mihalas} \&
  {Pradhan}}{{Seaton} et~al.}{1994}]{seaton94}
{Seaton} M.~J.,  {Yan} Y.,  {Mihalas} D.,    {Pradhan} A.~K.,  1994, \mnras,
  266, 805

\bibitem[\protect\citeauthoryear{{Simon}}{{Simon}}{1982}]{simon}
{Simon} N.~R.,  1982, \apjl, 260, L87

\bibitem[\protect\citeauthoryear{{Stankov} \& {Handler}}{{Stankov} \&
  {Handler}}{2005}]{handler}
{Stankov} A.,  {Handler} G.,  2005, \apjs, 158, 193

\bibitem[\protect\citeauthoryear{{Thoul}, {Aerts}, {Dupret}, {Scuflaire},
  {Korotin}, {Egorova}, {Andrievsky}, {Lehmann}, {Briquet}, {De Ridder} \&
  {Noels}}{{Thoul} et~al.}{2003}]{Thoul03}
{Thoul} A.,  {Aerts} C.,  {Dupret} M.~A.,  {Scuflaire} R.,  {Korotin} S.~A.,
  {Egorova} I.~A.,  {Andrievsky} S.~M.,  {Lehmann} H.,  {Briquet} M.,  {De
  Ridder} J.,    {Noels} A.,  2003, \aap, 406, 287

\bibitem[\protect\citeauthoryear{{Trundle}, {Dufton}, {Hunter}, {Evans},
  {Lennon}, {Smartt} \& {Ryans}}{{Trundle} et~al.}{2007}]{trundle07}
{Trundle} C.,  {Dufton} P.~L.,  {Hunter} I.,  {Evans} C.~J.,  {Lennon} D.~J.,
  {Smartt} S.~J.,    {Ryans} R.~S.~I.,  2007, \aap, 471, 625

\bibitem[\protect\citeauthoryear{{Turck-Chi{\`e}ze S. et
  al.,}}{{Turck-Chi{\`e}ze S. et al.,}}{2011}]{TC1}
{Turck-Chi{\`e}ze S. et al.,} 2011, Journal of Physics Conference Series, 271,
  012035

\bibitem[\protect\citeauthoryear{{Unno}, {Osaki}, {Ando}, {Saio} \&
  {Shibahashi}}{{Unno} et~al.}{1989}]{unno}
{Unno} W.,  {Osaki} Y.,  {Ando} H.,  {Saio} H.,    {Shibahashi} H.,  1989,
  {Nonradial oscillations of stars}

\bibitem[\protect\citeauthoryear{{Venn}}{{Venn}}{1999}]{venn99}
{Venn} K.~A.,  1999, \apj, 518, 405

\bibitem[\protect\citeauthoryear{{Venn} \& {Przybilla}}{{Venn} \&
  {Przybilla}}{2003}]{venn03bis}
{Venn} K.~A.,  {Przybilla} N.,  2003, in {C.~Charbonnel, D.~Schaerer, \&
  G.~Meynet} ed., Astronomical Society of the Pacific Conference Series
  Vol.~304 of Astronomical Society of the Pacific Conference Series, {New
  Nitrogen and Carbon in AF-supergiants (invited review)}.
p.~20

\bibitem[\protect\citeauthoryear{{Vermeij} \& {van der Hulst}}{{Vermeij} \&
  {van der Hulst}}{2002}]{vermeij02}
{Vermeij} R.,  {van der Hulst} J.~M.,  2002, \aap, 391, 1081

\bibitem[\protect\citeauthoryear{{Waelkens}}{{Waelkens}}{1991}]{waelkens}
{Waelkens} C.,  1991, \aap, 246, 453

\bibitem[\protect\citeauthoryear{{Zdravkov} \& {Pamyatnykh}}{{Zdravkov} \&
  {Pamyatnykh}}{2008}]{pami08}
{Zdravkov} T.,  {Pamyatnykh} A.~A.,  2008, Journal of Physics Conference
  Series, 118, 012079

\end{thebibliography}

\appendix

\section{SMC present-day abundances}

We describe below the available abundance data in the SMC (LMC is treated in Appendix B) and the criteria used to select what we regard as the most probable present-day value for each element. They are summarised in Table \ref{tab_abundances_smc} for the SMC. To ease readability, acronyms have been used in the following for the most cited references and are provided for convenience in Table \ref{tab_acronyms}. The absolute LTE abundances for B stars in NGC 330 of \citet*{lennon03} are in poor agreement with other studies and are not considered further.\\   

\begin{figure*}
\centering
\includegraphics[width=18.7cm]{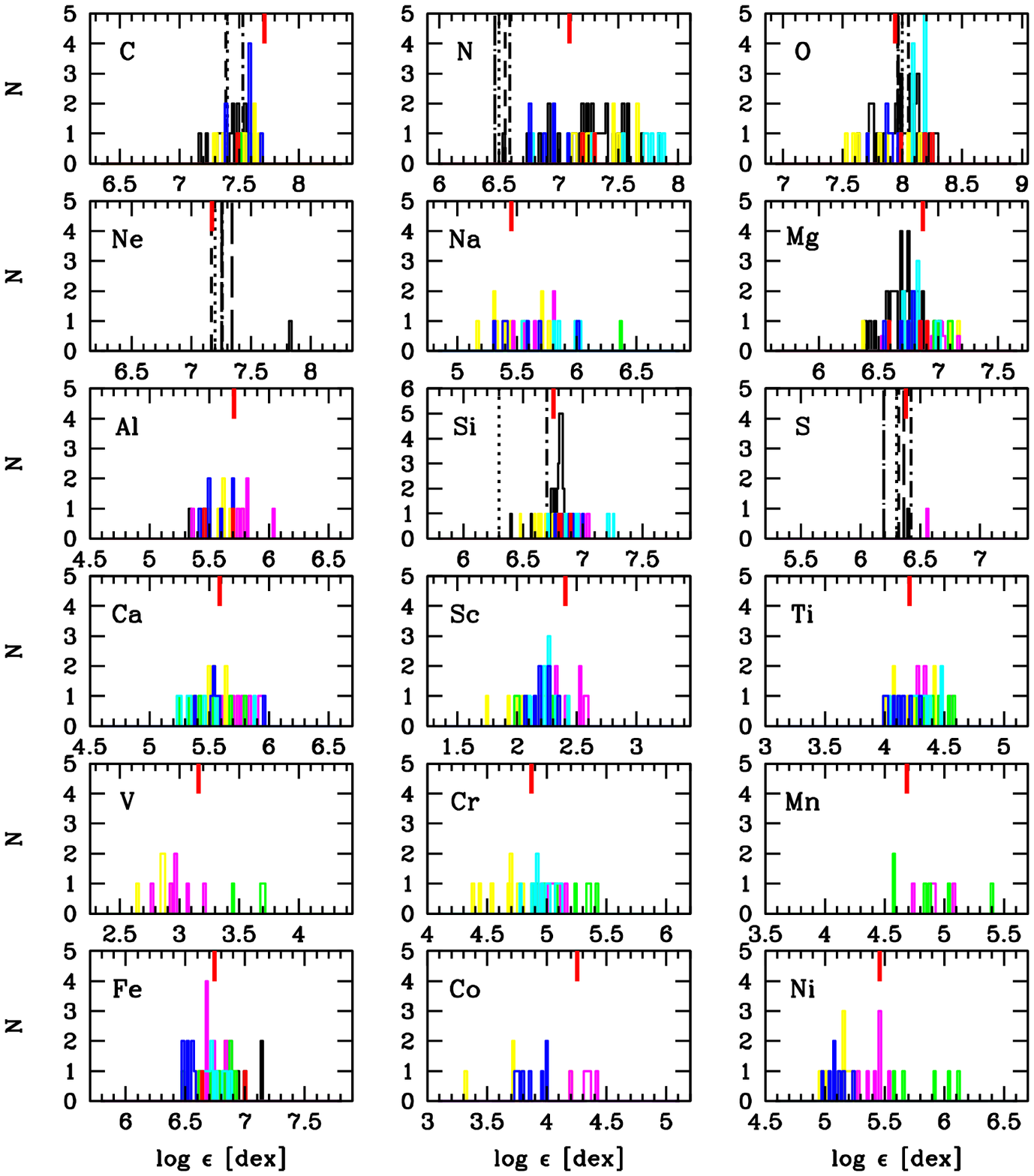}
\caption{SMC abundances. {\bf B stars} {\it black}: Hunter et al. (2005, 2007) and \citet{trundle07}, {\it red}: \citet{korn00} {\bf Cool supergiants} {\it magenta}: L98 and \citet{romaniello08}, {\it blue}: GW99, {\it cyan}: \citet{venn99} updated by \citet{venn03bis} for N, {\it green}: RB89, {\it yellow}: Hill et al. (1997) and Hill (1997, 1999) {\bf H II regions} {\it dotted line}: \citet{garnett99}, {\it short-dashed line}: \citet{reyes99}, {\it long-dashed line}: \citet{lebouteiller08}, {\it dot-short dashed line}: \citet{kurt99}, {\it dot-long dashed line}: \citet{vermeij02}. The vertical, red ticks at the top of each panel indicate the solar abundances of \citet{AGS09} -- 0.75 dex.}
\label{fig_abundances_smc}
\end{figure*}

\hspace*{-0.65cm} {\bf Carbon} Both \citet{korn00} and \citet{hunter05} suggest an NLTE value for B stars close to 7.5 dex. However, this estimate is only based on a handful of objects and must be cross checked against other sources. Despite the limitations affecting the carbon data from the VLT/FLAMES survey \citep{hunter07,trundle07}, see Sect. \ref{appendixb}, a close value is obtained from this much larger sample ($\sim$7.4 dex). On the other hand, all LTE studies of cool supergiants converge to a nearly identical abundance \nocite{hilletal97} \nocite{hill99}(Hill, Barbuy \& Spite 1997; Hill 1999; GW9; RB89). This is perhaps fortuitous, as one may expect some depletion because of mixing (albeit negative, the NLTE corections should, however, be small; A01). Equally pleasing is the good agreement with the data for \mbox{H\,{\sc ii}} regions when corrected to first order for dust depletion (Fig.~\ref{fig_abundances_smc}), suggesting that 7.5 dex is a robust estimate.\\   

\hspace*{-0.65cm} {\bf Nitrogen} Most B stars in the SMC show the products of CNO-cycle at their surface and are therefore N rich (as are a fortiori the cool supergiants). This is the case for the stars studied by \citet{korn00} and \citet{trundle07}. Data for \mbox{H\,{\sc ii}} regions cluster around 6.55 dex (Fig.~\ref{fig_abundances_smc}).
The consistency of these results and the negligible amount of nitrogen locked up in dust grains suggest that this estimate is reliable. Furthermore, it is identical to the value found for the B-type star AV304 by \citet{hunter05}. Several B stars observed during the VLT/FLAMES survey also have upper limits consistent with such a low value \citep{hunter07}.\\

\hspace*{-0.65cm} {\bf Oxygen} All NLTE studies of B stars are in good agreement and the mean value is adopted \citep{korn00,hunter05,hunter07,trundle07}. With the exception of Hill (1999), our best estimate is close to the values found for A, F and K supergiants (Hill et al. 1997; GW99; Venn 1999 [NLTE]). This is also the case for the data for \mbox{H\,{\sc ii}} regions (Fig.~\ref{fig_abundances_smc}).\\ 

\hspace*{-0.65cm} {\bf Neon} We base our abundance on observations of \mbox{H\,{\sc ii}} regions \citep{garnett99,reyes99,kurt99,lebouteiller08,vermeij02} that only show a small scatter. The LTE neon abundance derived for AV304 by \citet{hunter05} is obviously spurious (Fig.~\ref{fig_abundances_smc}).\\

\begin{table}
\centering
\caption{Acronyms used.}
\begin{tabular}{ll} \\\hline
Acronym & Reference\\
\hline 
A01 & \citet{andrievsky01}\\ 
GW99 & \citet{gonzalez99}\\
L98 & \citet{luck98}\\
RB89 & \citet{russel89}\\\hline
\end{tabular}
\label{tab_acronyms}
\end{table} 

\hspace*{-0.65cm} {\bf Sodium} There is no evidence for a strong Na enrichment arising from the dredge-up of Ne-Na cycled gas in SMC cool supergiants \citep{hill97}. We take the average of all the LTE values for yellow/red supergiants (Hill 1997, 1999\nocite{hill97}; Venn 1999\nocite{venn99}; L98; GW99), but omit the anomalously high abundance reported for AV127 by RB89.\\

\hspace*{-0.65cm} {\bf Magnesium} The numerous NLTE values in the literature for B stars indicate a value close to 6.75 dex without any discernable study-to-study differences \citep{korn00,hunter05,hunter07,trundle07}. The (mostly LTE) data for evolved objects suggest a mean  abundance $\sim$0.1 dex higher, which may not be significant (Hill 1997, 1999; L98; GW99; RB89; Venn 1999 [NLTE]).\\ 

\hspace*{-0.65cm} {\bf Aluminium} Only two B stars have NLTE abundances in the literature \citep{korn00}, but the mean value is indistinguishable from the abundances found for F and K supergiants (Hill 1997, 1999; GW99; L98). In contrast, the LTE abundance derived by \citet{hunter05} for AV304 is significantly lower.\\

\hspace*{-0.65cm} {\bf Silicon} We take the average of the NLTE values found for hot stars \citep{korn00,hunter05,hunter07,trundle07}, but once again note a very good overall agreement with the data for cool supergiants (Hill 1997, 1999; GW99; L98; Venn 1999).\\

\hspace*{-0.65cm} {\bf Sulphur} The adopted sulphur abundance is poorly constrained and only relies on the NLTE value measured by \citet{hunter05} in AV304. L98 obtained a LTE value $\sim$0.16 dex higher in one Cepheid, which is consistent within the errors. The (slightly) lower value inferred from \mbox{H\,{\sc ii}} regions (Fig.~\ref{fig_abundances_smc}) also provides some support for our estimate, as some dust depletion is expected.\\
 
\hspace*{-0.65cm} {\bf Argon} All studies of the ionised gas in \mbox{H\,{\sc ii}} regions yield very similar results and we assume the mean value here \citep{garnett99,reyes99,kurt99,lebouteiller08,vermeij02}. \\
  
\hspace*{-0.65cm} {\bf Calcium, Scandium and Titanium} We take the median of the abundances derived for cool stars (Hill 1997, 1999; GW99; L98; RB89; Venn 1999).\\

\hspace*{-0.65cm} {\bf Vanadium} There is a large spread in the published values (Fig.~\ref{fig_abundances_smc}). We average the values of Hill (1999), L98 and RB89. The abundance of this element is uncertain.\\

\hspace*{-0.65cm} {\bf Chromium} We consider the typical value found for cool supergiants (Hill 1997, 1999; L98; RB89; Venn 1999).\\

\hspace*{-0.65cm} {\bf Manganese} The values found by L98 and RB89 have been used.\\
  
\hspace*{-0.65cm} {\bf Iron} Compared to the LMC, much fewer B-type stars have had their iron content determined and the scatter from one study to another is also slightly larger: Hunter et al. (2005; 6.63 dex in LTE), Korn et al. (2000; 6.82 dex in NLTE) and Trundle et al. (2007; 6.93 dex in LTE). We prefer to rely on the numerous observations of yellow/red supergiants that suggest an abundance $\sim$6.75 dex (Hill 1997, 1999; L98; RB89; Venn 1999; Romaniello et al. 2008\nocite{romaniello08}), but note that the NLTE value of \citet{korn00} for two B stars is only 0.07 dex away from this value.\\

\hspace*{-0.65cm} {\bf Cobalt} We use the values of Hill (1999), L98 and GW99, but omit the star with an anomalously low abundance in the sample of Hill (1999).\\

\hspace*{-0.65cm} {\bf Nickel} The values of Hill (1997, 1999), GW99, L98 and RB89 have been used.\\

\section{LMC present-day abundances}
\label{appendixb}
As for the SMC, we present here a compilation of abundances for the LMC and the criteria used to select what are the most probable present-day value for each element. They are summarised in Table \ref{tab_abundances_lmc}. Acronyms of Table \ref{tab_acronyms} are still used in this section.\\

\begin{figure*}
\centering
\includegraphics[width=18.7cm]{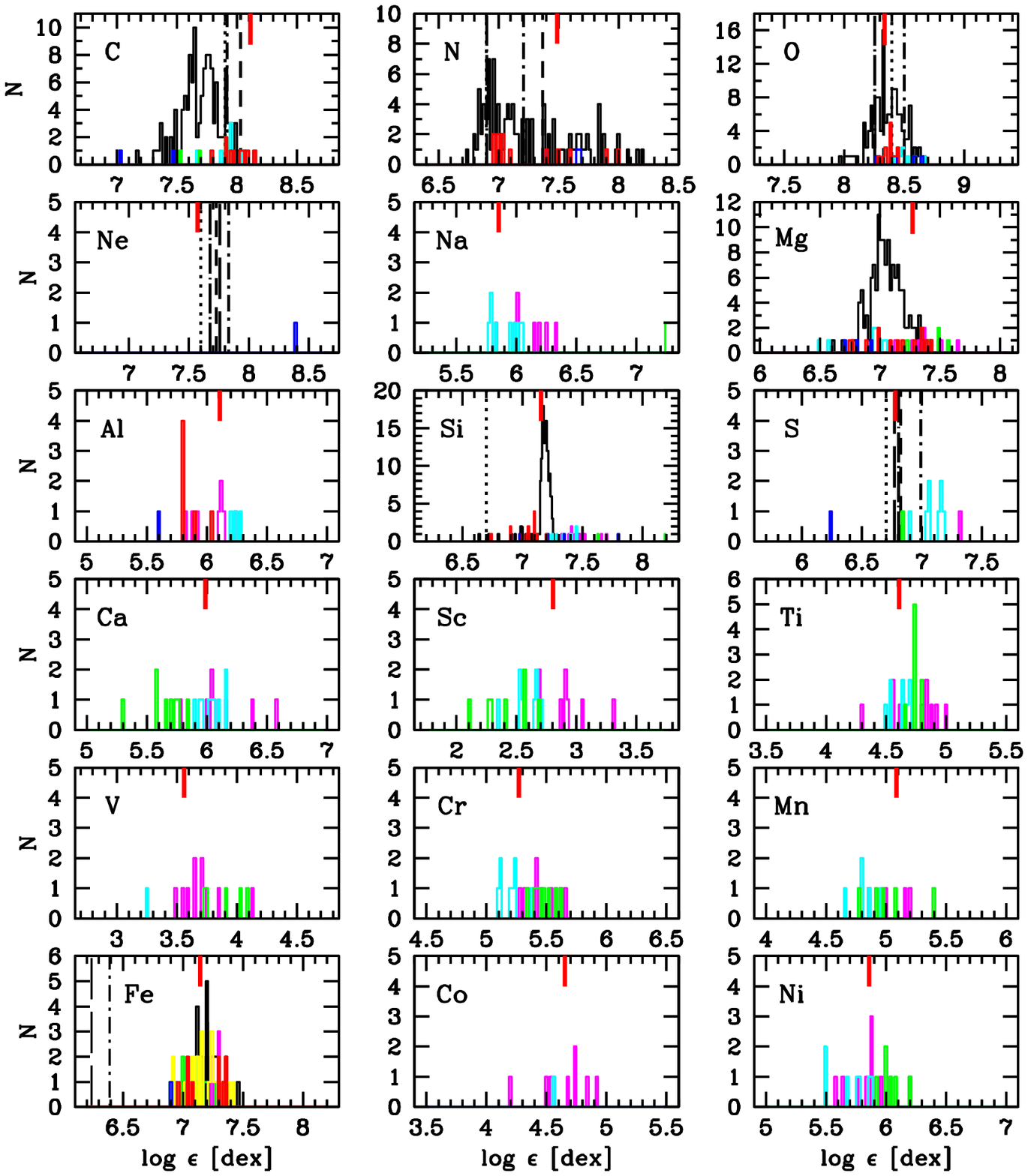}
\caption{LMC abundances. {\bf B stars} {\it black}: \citet{hunter08} and \citet{trundle07}, {\it blue}: \citet{rolleston}, {\it red}: Korn et al. (2000, 2002, 2005) {\bf Cool supergiants} {\it magenta}: L98, {\it cyan}: A01, {\it green}: RB89, {\it yellow}: \citet{romaniello08} {\bf H II regions} {\it dotted line}: \citet{garnett99}, {\it short-dashed line}: \citet{reyes99}, {\it long-dashed line}: \citet{lebouteiller08}, {\it dot-short dashed line}: \citet{peimbert03}, {\it dot-long dashed line}: \citet{vermeij02}. The vertical, red ticks at the top of each panel indicate the solar abundances of \citet{AGS09} -- 0.35 dex.}
\label{fig_abundances_lmc}
\end{figure*}

{\bf Carbon} We adopt the NLTE abundances derived by Korn et al. (2000, 2002, 2005). Carbon abundances are available for the large number of stars observed in the course of the VLT/FLAMES survey of massive stars \citep{hunter08}. However, because these results are mainly based on lines of which reliability is doubtful (e.g. \mbox{C\,{\sc ii}} $\lambda$4267) we do not consider these results here. The few LTE abundances of \citet*{rolleston} are in poor agreement with other studies. The slightly lower abundances found for cool supergiants ($\sim$0.15 dex; RB89; A01 [NLTE]), if physical, might be explained by 
 evolutionary effects. The values for \mbox{H\,{\sc ii}} regions are in very good agreement 
 (Fig.~\ref{fig_abundances_lmc}). Only a small amount of carbon is believed to be trapped in dust grains ($\sim$0.1 dex in the Orion nebula according to Esteban et al. 1998\nocite{esteban98}).\\   

\hspace*{-0.65cm} {\bf Nitrogen} Nitrogen abundances show a very wide spread in B stars in the LMC, with most of them exhibiting a strong excess as a result of mixing. We adopt the baseline abundances of \citet{hunter08} corresponding to the lowest values found in their sample. This is consistent with the values of Korn et al. (2000, 2002, 2005). The stars analysed by \citet{rolleston} are clearly enriched. The data for \mbox{H\,{\sc ii}} regions show a large scatter (Fig.~\ref{fig_abundances_lmc}), but our estimate is in excellent agreement with \citet{garnett99} and \citet{vermeij02}. Nitrogen is not expected to be significantly depleted in the gas phase.\\

\hspace*{-0.65cm} {\bf Oxygen} The oxygen abundance is taken from \citet{hunter08} and is in very good agreement with the values of Rolleston et al. (2002; LTE) and Korn et al. (2000, 2002, 2005; NLTE). This value is reasonably consistent with the NLTE values of A01 for F supergiants ($\sim$8.5 dex) and the results for \mbox{H\,{\sc ii}} regions (Fig.~\ref{fig_abundances_lmc}). Only a small amount of oxygen is believed to be trapped in dust grains \citep[$\sim$0.08 dex in the Orion nebula according to][]{esteban98}.\\ 

\hspace*{-0.65cm} {\bf Neon} We base our abundance on observations of \mbox{H\,{\sc ii}} regions \citep{garnett99,reyes99,lebouteiller08,peimbert03,vermeij02}. There is a good concordance between these values and, as a noble gas, neon is not embedded in dust grains. \citet{rolleston96} derived the LTE neon abundance in LH 104--24, but this value is discrepant and very likely grossly overestimated (Fig.~\ref{fig_abundances_lmc}).\\
 
\hspace*{-0.65cm} {\bf Sodium} There is no evidence for a strong Na enrichment arising from convective mixing in LMC cool supergiants (A01). We adopt the mean NLTE abundance of A01 for F supergiants, which compares favourably with the values of L98 for Cepheids if NLTE corrections of the order of --0.1 dex are applied (A01). The value for one star of RB89 is discrepant.\\

\hspace*{-0.65cm} {\bf Magnesium} We choose the mean value of 7.05 dex found by \citet{hunter08}. Other studies based on smaller samples of early-type stars give very similar results \citep{korn00,korn02,korn05,rolleston}. Mean LTE abundances between 6.88 and 7.44 dex have been reported for cool supergiants (A01; L98; RB89).\\ 

\hspace*{-0.65cm} {\bf Aluminium} The LTE value of \citet{rolleston96} for PS34-16 is on the low side (Fig.~\ref{fig_abundances_lmc}). We adopt the NLTE abundances of \citet{korn00}. Higher values are suggested from LTE studies of F supergiants ($\sim$6.1 dex; A01; L98).\\

\hspace*{-0.65cm} {\bf Silicon} The Si abundances of a large sample of B stars have been derived in the framework of the VLT/FLAMES survey \citep{hunter08} and we use these values here. This mean value is in agreement with the LTE estimates of \citet{rolleston}, while it lies in between the results for other B stars in NLTE \citep{korn00,korn02,korn05} and red supergiants in LTE (A01; L98; RB89). The abundances from \mbox{H\,{\sc ii}} regions are unreliable because of severe dust depletion. \\

\hspace*{-0.65cm} {\bf Sulphur} The LTE abundance estimated by \citet{rolleston} clearly appears underestimated (Fig.~\ref{fig_abundances_lmc}). We adopt the mean of the few LTE values reported for evolved objects (A01; L98; RB89). The significantly lower abundance found from \mbox{H\,{\sc ii}} regions likely arises from dust depletion \citep[e.g.][]{lebouteiller08}.\\
 
\hspace*{-0.65cm} {\bf Argon} No stellar studies exist and we base our results on observations of \mbox{H\,{\sc ii}} regions, which are in close agreement \citep{garnett99,reyes99,lebouteiller08,peimbert03,vermeij02}.\\
  
\hspace*{-0.65cm} {\bf Calcium} We only use the values of A01 and L98, as the values of RB89 based on lower quality data seem slightly discrepant (Fig.~\ref{fig_abundances_lmc}).\\

\hspace*{-0.65cm} {\bf Scandium} There is a large spread in the results of the different studies (A01; L98; RB89), but we take the mean as a representative value.\\

\hspace*{-0.65cm} {\bf Titanium} Excellent agreement is found between the various studies (Fig.~\ref{fig_abundances_lmc}) and we adopt the mean value (A01; L98; RB89).\\

\hspace*{-0.65cm} {\bf Vanadium} We average the values of L98 and RB89 and omit the discrepant value found for one star by A01 (Fig.~\ref{fig_abundances_lmc}).\\

\hspace*{-0.65cm} {\bf Chromium and Manganese} A01 find on average slightly lower values than L98 and RB89 for both elements. As the distributions overlap, however, we simply take the mean value.\\

\hspace*{-0.65cm} {\bf Iron} We ignore the LTE value of \citet{rolleston} for one star based on 2 lines differing by up to 0.6 dex. The mean values for the other B star studies are in fairly good agreement: Korn et al. (2000; 7.09 dex in NLTE), Trundle et al. (2007; 7.21 dex in LTE) and Korn et al. (2002; 7.32 dex in LTE). Considering the smallness of the NLTE corrections \citep{korn00}, we can compute the grand average of all these values to obtain $\sim$7.21 dex. On the other hand, the LTE analyses of the F-type supergiants suggest $\sim$7.14 dex (A01; L98; RB89; Romaniello et al. 08). We assume 7.15 dex as a compromise. Iron is dramatically depleted in dust grains, such that \mbox{H\,{\sc ii}} regions do not offer any constraints (Fig.~\ref{fig_abundances_lmc}).\\

\hspace*{-0.65cm} {\bf Cobalt} We take the mean of the values of A01 and L98.\\

\hspace*{-0.65cm} {\bf Nickel} We take the mean of the values of A01, L98 and RB89.\\

\end{document}